\documentclass{aa}  
\usepackage[utf8]{inputenc}
\usepackage{booktabs}

\usepackage{graphicx}
\usepackage[caption=false]{subfig}
\usepackage{verbatim}
\usepackage{txfonts}
\usepackage{xcolor}
\newcommand{\maxi}{MAXI~J1957+032}

\newcommand{\nicer}{NICER}
\newcommand{\xmm}{XMM-Newton}

\newcommand{\nustar}{NuSTAR}

\newcommand{\swift}{Swift}
\newcommand{\swiftxrt}{Swift/XRT}

\newcommand{\maxigsc}{MAXI/GSC}

\newif\ifrefchanges
\refchangesfalse   

\newcommand{\rev}[1]{%
  \ifrefchanges
    \textbf{#1}%
  \else
    #1%
  \fi
}

\DeclareUnicodeCharacter{2212}{-}
\begin{document}

   \title{Flashing fast: characterising the 2025 outburst of \maxi{}}
   \titlerunning{The 2025 outburst of \maxi{}}
    \authorrunning{A.~Sanna et al.}

\author{A.~Sanna\inst{1}
        \and G.~Illiano\inst{2}
        \and M.~C.~Baglio\inst{2}
        \and D.~M.~Russell\inst{3}
        \and A.~Borghese\inst{4}
        \and A.~Miraval~Zanon\inst{5}
        \and A.~Marino\inst{6,7,8}
        \and A.~Riggio\inst{1}
        \and A.~Papitto\inst{9}
        \and K.~Alabarta\inst{2}
        \and T.~Di~Salvo\inst{10}
        \and A.~Anitra\inst{1,10}
        \and L.~Burderi\inst{1,8}
        \and F.~Lewis\inst{11,12}
        \and R.~Iaria\inst{10}
        \and D.~A.~H.~Buckley\inst{13}
        }
\institute{
  Dipartimento di Fisica, Universit\`a degli Studi di Cagliari, SP
Monserrato-Sestu, KM 0.7, Monserrato, 09042 Italy \\
  \email{andrea.sanna@dsf.unica.it}
  \and
  INAF-Osservatorio Astronomico di Brera, Via Bianchi 46, I-23807, Merate (LC), Italy
  \and
  Center for Astrophysics and Space Science (CASS), New York University Abu Dhabi, PO Box 129188, Abu Dhabi, UAE
  \and
  European Space Science (ESA), European Space Astronomy Center (ESAC), Camino Bajo del Castillo s/n, E-28692 Villanueva de la Ca\~{n}ada, Madrid
  \and
  ASI - Agenzia Spaziale Italiana, Via del Politecnico snc, 00133 Roma, Italy
  \and
  Institute of Space Sciences (ICE, CSIC), Campus UAB, Carrer de Can Magrans s/n, E-08193 Barcelona, Spain
  \and
  Institut d'Estudis Espacials de Catalunya (IEEC), 08860 Castelldefels (Barcelona), Spain
  \and
  INAF-IASF Palermo, Via Ugo La Malfa 153, 90146 Palermo, Italy
  \and 
  INAF-Osservatorio Astronomico di Roma, Via Frascati 33, I-00076 Monte Porzio Catone, (RM), Italy
  \and
  Dipartimento di Fisica e Chimica - Emilio Segre, Università di Palermo, via Archirafi 36 - 90123 Palermo, Italy
  \and
  Faulkes Telescope Project, School of Physics and Astronomy, Cardiff University, The Parade, Cardiff, CF24 3AA, Wales, UK
  \and
  The Schools’ Observatory, Astrophysics Research Institute, Liverpool John Moores University, 146 Brownlow Hill, Liverpool L3 5RF, UK
  \and
  South African Astronomical Observatory, P.O Box 9, Observatory, 7935 Cape Town, South Africa
}

   \date{Received 27 October 2025; accepted 11 December 2025 }

\abstract{\maxi{} is an accreting millisecond X-ray pulsar showing brief, recurrent outbursts within an ultra-compact $\approx1$~h orbit.}{We characterise X-ray timing, spectral, and optical properties during the 2025 outburst and measure the long-term spin evolution relative to its previous 2022 outburst.}{We analysed X-ray observations from \xmm{}, \swift{}, and \nustar{}, together with contemporaneous optical photometry obtained with LCO during the 2025 outburst.
X-ray timing analysis included standard epoch-folding and coherent searches, while energy-resolved pulse profiles were studied through harmonic decomposition.
Spectral fits used absorbed thermal–Comptonisation models complemented by a soft blackbody component, whose emission radius suggests it likely originates from a hotspot on the neutron star surface.}{Coherent pulsations were detected at $\nu \approx 313.6$~Hz, with no measurable frequency derivative within the \xmm{} exposure.
By comparing with the 2022 outburst, we find a long-term spin-down of $\langle\dot\nu\rangle \sim -2\times10^{-14}$~Hz~s$^{-1}$, consistent with magnetic dipole braking during quiescence.
The pulse shape is almost sinusoidal, showing significant power at the fundamental, second, and fifth harmonics.
The fractional amplitude decreases with increasing flux and exhibits soft lags extending to a few keV.
The X-ray spectrum between 0.5 and 10~keV is well reproduced by a thermal–Comptonised continuum with photon index $\Gamma \approx 2.4$, plus a cool blackbody with $kT \approx 0.23$~keV.
No reflection or Fe-line features are detected.
Assuming $R_{\rm m} \lesssim R_{\rm co}$, the magnetic field is limited to $B_{\rm s} \approx (0.5$–$3)\times10^{8}$~G for $d = (5 \pm 2)$~kpc and truncation factor $\xi = 0.3$–$0.5$, \rev{lower than the upper limit implied by the secular spin-down ($B_{\rm p}\lesssim10^{9}$~G)}, possibly indicating a mildly leaky propeller regime.
The optical emission follows the neutron-star branch of the $L_{\mathrm{OIR}}$–$L_X$ relation, consistent with X-ray reprocessing in a compact accretion disc.
The optical SEDs are broadly flat, supporting irradiation-dominated disc emission, while an early red excess suggests a jet contribution during the initial hard X-ray phase.
A delayed optical peak relative to the X-rays may reflect the outward propagation of a heating front through the disc, consistent with rapid disc evolution in short-lived outbursts.}{}

   \keywords{accretion -- neutron stars -- Pulsars: general -- X-rays: binaries -- X-rays: individuals: MAXI~J1957$+$032
               }

\maketitle
%

\section{Introduction}

Neutron stars (NSs) hosted in low-mass X-ray binaries (LMXBs) owe their typical rapid, i.e., of the order of millisecond, periods to a prolonged accretion of angular momentum from a companion that barely reaches the solar-mass scale \citep{Alpar82, Wijnands:1998vk}. Gas streaming through the inner Lagrange point settles into a Keplerian disc \rev{\citep[e.g.][]{Shakura73, Frank02}}. Once the flow is magnetically channelled onto the NS surface \rev{\citep{Ghosh:1979aa, Ghosh:1979ws}}, hotspots sweep across our line of sight, imprinting X-ray pulsations at hundreds of hertz \rev{\citep[see, e.g.,][]{Psaltis99, Poutanen:2006aa}}.  These "recycled" accretors, first predicted four decades ago and observed soon after, now number a few dozen \citep[see][for extensive reviews]{Di-Salvo:2020va, Patruno:2021vs} and bridge the evolutionary gap between \rev{non-pulsating LMXBs (in which no coherent millisecond pulsations have been detected so far)} and radio millisecond pulsars.

A crucial trait of the class of accreting millisecond X-ray pulsars (AMXPs) is their transience \rev{\citep[see, e.g.,][]{Wijnands:1998vk, Campana2018a}}. A possible scenario envisions a disc that is usually too cool and tenuous to conduct matter inward \rev{\citep{Lasota01}}. Still, every so often, a thermal-viscous runaway may heat the flow, raise the accretion rate by orders of magnitude and ignite an X-ray outburst. These events are frequently accompanied by optical and near-infrared emission driven by X-ray irradiation of the accretion disc and, in some cases, of the companion star \rev{\citep[see, e.g.,][]{vanparadijs:1994aa, hynes:2005aa, russel:2006aa, russel:2007aa}}.
Monitoring this component provides a complementary view of the accretion process, as the optical flux reflects both the geometry and the efficiency of X-ray reprocessing.
In the canonical disc-instability picture \citep[][]{Meyer:1981aa, King:1998aa, Hameury:2020aa}, outbursts can last for weeks to months, \rev{yet theory also predicts markedly briefer episodes if the accretion disc has a particularly small radial extent, as anticipated for ultra-compact systems \citep[see, e.g.,][]{Hameury:2016aa, Marino:2019vq}, or if it is truncated close to the magnetosphere \citep[see, e.g.,][]{Burderi98b, Kulkarni:2013tf, Bozzo:2018aa}}. Simulations tailored to hydrogen-poor predictions for ultra-compact binaries suggest that eruptive cycles can empty the entire reservoir in only a few days, an expectation supported by observations of several AMXPs \citep[see, e.g.,][]{Hameury:2016aa, Marino:2019vq, heinke:2025aa}. Statistical work indicates that the outburst recurrence rate, and to a lesser extent the duration-depends on the orbital period, reflecting differences in disc size and mass transfer across LMXBs. In particular, systems with $P_{\rm orb}\lesssim 12$~h tend to show much lower outburst rates than longer-period systems \citep{Lin:2019aa}. However, this trend does not appear to hold within the subclass of AMXPs, which occupy the short-period regime (typically $P_{\rm orb} \lesssim 5$~h) and show no clear correlation between orbital period and outburst frequency. Their uniformly compact accretion discs and low mass-transfer rates likely produce infrequent and irregular outbursts, making them a distinct \rev{observational} population within the broader LMXB family. \rev{Although AMXPs are not generally classified as Very-Faint X-ray Transients (VFXTs), their low outburst luminosities overlap the $L_X\sim10^{34}$--$10^{36}$~erg~s$^{-1}$ regime where NS VFXTs show similar softening trends \citep[see, e.g.,][]{Wijnands:2015aa}.}

\begin{figure}
    \centering
    \includegraphics[width=0.4\textwidth]{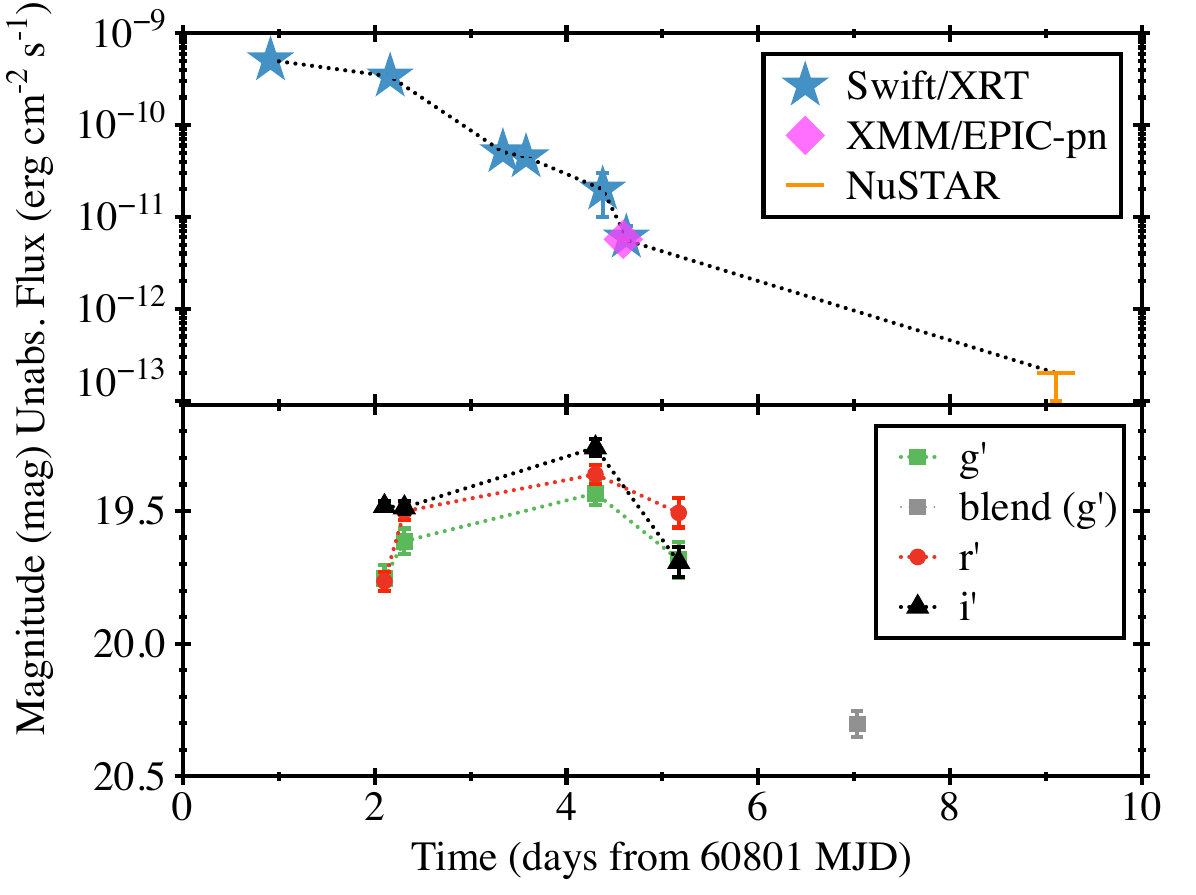}
    \caption{\rev{Multi-wavelength light curves of \maxi{} during the 2025 outburst. The upper panel shows the 0.5--10~keV unabsorbed flux measured with \swift{}/XRT (blue points) and the \xmm{}/EPIC-pn observation (pink point). The lower panel shows the optical $g'$, $r'$, and $i'$ magnitudes from LCO (see Sect.~\ref{sec:optical_data}). An additional observation per-
    formed on May 13, 2025 (MJD 60808), corresponding to the
    blend of \maxi{} with a $g'\sim20$ nearby star, is shown
    for comparison. Error bars represent 1$\sigma$ uncertainties.}}
    \label{fig:xray_opt_lc}
\end{figure}

\maxi{} sits at the sharp end of this "flash-outburst" sub-population.  Since its discovery \citep{Negoro:2015vw}, the source has \rev{undergone roughly half a dozen outbursts; each outburst has risen, peaked at $\sim10^{36}$~erg~s$^{-1}$ (0.5--10~keV), and faded back into quiescence in $\lesssim 5$~d} at flux values of a few $10^{33}$~erg~s$^{-1}$ \citep[see, e.g.,][and references therein]{Ravi:2017tl, Mata-Sanchez:2017vl}. Coherent pulsations at $\sim$314 Hz and a Doppler-shifted orbital period of $\simeq$1~h certify the system as an ultra-compact accreting millisecond pulsar, implying a disc radius of only a few $10^9$~cm and an extremely \rev{low-mass} donor, possibly a brown dwarf \citep{Ng:2022uc, Sanna:2022vi} or a carbon-oxygen white dwarf \citep{Ravi:2017tl}. The latest outburst detected by the Einstein Probe telescope on 2025 May 6 \citep{Sun:2025aa}, and confirmed by \maxigsc{} and \swiftxrt{} \citep{Negoro:2025aa, Williams_2025ATel17172_Swift, Illiano_2025ATel17187} once again followed the same rapid template, barely reaching a few $10^{36}$~erg~s$^{-1}$ before disappearing below monitor detectability within a week \citep{Li:2025aa}. Timing analysis of the Einstein Probe dataset confirmed a new outburst of the source by detecting coherent pulsations at the expected spin frequency reported in literature \citep{Li:2025aa}. \rev{A broad-band view of the 2025 outburst, combining the X-ray and optical coverage, is shown in Fig.~\ref{fig:xray_opt_lc}.}

Here, we report on the timing and spectral properties of \maxi{} as observed with \xmm{} during a dedicated Target of Opportunity observation in its latest outburst.
To trace the spectral evolution, we also utilise a \textit{Swift}/XRT monitoring campaign, and we include optical photometry obtained during the same epoch to investigate the behaviour of the irradiated accretion disk. \rev{In addition to the \xmm{} and \swift{} datasets, a \nustar{} observation was obtained nine days after the outburst peak. Since the source was not significantly detected in this pointing and no useful timing or spectral constraints could be derived, the dataset is not used in our analysis. For completeness, the observation and basic count-rate estimates are described in Sec.~\ref{sec:nustar}.}


\section{Observation and data reduction}

\subsection{\xmm{}}

\xmm{} \citep{Jansen2001} performed a target of opportunity observation of \maxi{} (Obs. ID. 0971190201) on 2025 May 10, starting from 08:43 UTC till 22:02 UTC. Different instruments were set up for the observations, including the EPIC-pn (PN) camera, which was operated in timing mode for the first $\sim 32.8$~ks and switched to burst mode for the remaining $\sim 8.9$~ks of coverage. Both EPIC-MOS (1--2) cameras were operated in timing mode, while the RGS was in spectroscopy mode. Although both EPIC-MOS cameras were configured in Timing mode, we did not use MOS data for timing analysis due to limited calibration \citep[see, e.g.,][]{valencic:2016aa} and absolute-timing scatter of the order of $\approx \pm 10$~ms (estimates from Crab cross-checks). For this work, we focused solely on the PN dataset taken in timing mode, discarding the burst-mode interval because its 3\% duty cycle resulted in only $\sim1000$ surviving photons, too few for reliable spectral or timing analysis. 

We performed screening and cleaning of the PN events using the Science Analysis Software (SAS) v.21 with up-to-date calibration files. We extracted source photons limiting their energy in the range 0.5–10 keV, and by retaining only calibrated photons characterised by \textsc{PATTERN $\leq$ 4} and \textsc{(FLAG = 0)}. We isolated source and background photons from a 23-pixel-wide strip (RAWX=26–48) centred on the brightest pixel column and an 11-pixel-wide (RAWX=3–13) strip in the tail of the RAWX distribution, respectively. We investigated the presence of flaring activity above 10 keV by generating a 20-s time-bin-resolution light curve from PN timing-mode events extracted from the source, and we identified none. From the same dataset, we generated the 0.5–10 keV light curve, showing an average count rate of around 3 counts/s, with a clear decreasing trend from $\sim 4$ to $\sim 1.5$ counts/s. We did not observe either type-I X-ray burst or partial occultation of the source count rate during the PN exposure time. For the timing analysis, we applied barycentric corrections of the photon arrival times to the Solar System barycentre by using the \textsc{barycen} tool (DE-405 solar system ephemeris) and adopting the best available source coordinates of the source \citep{Chakrabarty:2016vp} reported in Table~\ref{table:timing}.  

\rev{The PN light curve described in this section is shown, together with the optical monitoring, in the multiwavelength outburst overview presented in Fig.~\ref{fig:xray_opt_lc}.}

\subsection{\nustar{}}\label{sec:nustar}
\nustar{} triggered a pointed observation of \maxi{} (Obs.ID. 91101312002)
starting on May 14, 2025, 13:45 UTC and ending on May 15, 2025, 15:25 UTC for a total coverage time of $\sim$92.4~ks.
We processed the \nustar{} observation using the standard pipeline in \textsc{NUSTARDAS} (HEASOFT v6.33.2), applying the default screening and filtering to produce cleaned event lists for FPMA and FPMB. Source and background events were extracted with circular apertures centred on the target and on a source-free region in the same detector quadrant, respectively; light curves and spectra (with response files) were generated with \textsc{nuproducts}. Background-subtracted light curves for each module and the combined dataset were obtained with \textsc{lcmath}.

The exposure after screening amounts to $\sim$46~ks. In the whole 3–80~keV band, we measure a total of 2779 \emph{net} counts when summing FPMA and FPMB, corresponding to a mean background-subtracted rate of $\approx6.1\times10^{-2}$~counts/s (i.e., $\sim3\times10^{-2}$~counts/s per module). 

\rev{Using the spectral parameters from the nearest \swiftxrt{} observation, we estimated the corresponding \nustar{} 3–20~keV flux via WebPIMMS. The inferred flux, of order a few $\times 10^{-13}$~erg~cm$^{-2}$~s$^{-1}$, is shown in Fig.~\ref{fig:xray_opt_lc}, and confirms that \maxi{} had already returned to quiescence, providing no useful timing or spectral constraints.}

\subsection{\swift{}} \label{sec:swift_data}

The \swift{} \citep{Gehrels_2004ApJ} X-ray Telescope (XRT; \citealt{Burrows_2005SSRv}) observed \maxi{} ten times during its 2025 outburst (see Table~\ref{tab:swift_obs}). The first XRT observation was carried out in photon counting (PC) mode on 2025 May 6 \citep{Williams_2025ATel17172_Swift}, confirming the onset of the outburst. A dedicated monitoring campaign then followed, with a series of observations performed approximately every six hours in windowed timing (WT) mode between 2025 May 8 and May 11 (PI: Illiano), tracking the outburst evolution with high cadence. Finally, a last observation in PC mode was conducted on 2025 May 15 to confirm the source's return to quiescence \citep{Illiano_2025ATel17187}.

The raw, level-1 data were processed using \texttt{xrtpipeline} with standard parameters. The first PC-mode observation (ObsID: 00019768001) was affected by pile-up. To mitigate it, we extracted source photons from an annular region centred on the source coordinates with inner and outer radii of 16 and 30 pixels, respectively (1 pixel = 2.36 arcsec). The background was estimated from a concentric annular region with inner and outer radii of 40 and 80 pixels, respectively. The subsequent set of observations, carried out in WT mode, was analysed using a circular region with a radius of 20 pixels, centred on the source position for extracting the source photons, and a background region of the same size located away from the source.

All spectra were extracted in the 0.3–10~keV energy range and rebinned to ensure a minimum of 25 counts per bin, except for two WT-mode observations (ObsIDs: 00033770039 and 00033770040), which had low statistics and a background-dominated high-energy range. These spectra were instead grouped to have at least 10 counts per bin and fitted using a modified version of the Cash statistic that accounts for background effects (W-statistic\footnote{see \url{https://heasarc.gsfc.nasa.gov/docs/software/xspec/manual/node119.html}}) in the 0.5–10~keV and 0.5–9~keV energy ranges, respectively.

The spectra from ObsIDs 00033770041 and 00033770042 were excluded from the analysis because the source was detected only below 3 keV in both cases, which prevented a reliable determination of spectral parameters. We also attempted to combine these two observations to improve the statistics, but the resulting renormalisation factors used in the spectral modelling showed inconsistencies, as expected given the rapid decay phase of the source \rev{(see Fig.~\ref{fig:xray_opt_lc}, top panel)}. Finally, the last PC-mode observation was not analysed, as the source was no longer detected, consistent with the quiescent state reported by \citet{Illiano_2025ATel17187}. Both the \texttt{sosta} tool and the \swiftxrt{} online product generator returned a 3$\sigma$ upper limit on the count rate of $\sim$$6 \times 10^{-3}$~counts/s in $\sim$1.8 ks of exposure. Assuming the latest spectrum reported in Table~\ref{Table:spectra_Swift}, this corresponds to a 3$\sigma$ upper limit on the unabsorbed 0.5–10~keV flux of $\sim$$2 \times 10^{-13}$~erg~cm$^{-2}$~s$^{-1}$. 

\rev{The time evolution of the XRT unabsorbed fluxes is shown in Fig.~\ref{fig:xray_opt_lc}.}

\subsection{Las Cumbres Observatory}\label{sec:optical_data}

\maxi{} was monitored during its 2025 outburst in the optical band using the 1~m and 2~m telescopes of the Las Cumbres Observatory (LCO) network, as part of an ongoing program monitoring approximately 50 LMXBs \citep{Lewis2008}. Observations were obtained with the SDSS $g'$, $r'$, $i'$, and $z$ filters, beginning on 2025 May 6 (MJD~60803), approximately one day after the first X-ray detection, and continuing until 2025 May 30. The target lies within $\sim$2'' of a nearby star with $g'\sim20$~mag 
Depending on the seeing conditions and on the brightness of \maxi{} at the time of observation, the two sources could appear partially blended. When the transient was bright, close to the outburst peak, it could be clearly resolved and accurately measured. At fainter stages, however, especially under poor seeing, the two objects became indistinguishable, and all such images were identified through visual inspection and discarded from the analysis. Given the short duration of the outburst, \maxi{} became too faint to be clearly separated from the nearby star after only about three days of optical monitoring. This work will therefore focus mainly on optical observations acquired between 2025 May 8 and May 11 (MJD~60803–60806). All observations during this period were obtained with the 1~m telescopes of the LCO network.

The $z$-band frames were collected only when the source was blended or too faint, while $Y$-band data yielded upper limits during unblended epochs. Accordingly, we restrict our analysis to the $g’$, $r’$, and $i’$ bands. Photometry was extracted with the \textsc{XB-NEWS} pipeline \citep{Russell2019, Goodwin2020}, which performs multi-aperture photometry \citep[MAP;][]{Stetson90} and automated astrometric and photometric calibration. Details of the reduction, calibration procedure, and final magnitudes are provided in Appendix \ref{app:optical} (see Table \ref{tab:optical_photometry}). 

\rev{The resulting optical light curves are shown together with the X-ray evolution in Fig.~\ref{fig:xray_opt_lc}.}

\section{Results}

\subsection{Timing analysis} \label{sec:timing_analysis}

All photon arrival times from the PN camera were barycentered using \textsc{barycen} and corrected for the binary motion assuming a circular orbit. We propagated the orbital solution from the 2022 \nicer{} outburst \citep{Sanna:2022vi} to the present epoch and scanned the time of ascending node within its $\pm3\sigma$ range in 1-s steps.

\rev{For each trial, photon arrival times were first corrected to the Solar System 
barycentre using \textsc{barycen}, which removes the delays caused by the 
Earth’s motion. We then applied an orbital demodulation by subtracting the 
Roemer delay expected from a circular binary orbit. After this two–step 
procedure, the events were folded into 32 phase bins while scanning the spin frequency around the 2022 
value} ($\nu_0=313.64374049$~Hz) with $10^{-7}$~Hz increments. The $\chi^2$ periodogram yielded a single significant maximum at $T_{\mathrm{asc}}=60805.36786(1)$~MJD and $\nu=313.64373842(35)$~Hz. The frequency uncertainty was estimated from the curvature of the $\chi^2(\nu)$ peak and via bootstrap resampling of the photon list, yielding consistent results within \rev{2\%}.

Using the optimal orbital solution and spin frequency from the folding search, we conducted a fully phase-coherent analysis of the PN events. To avoid mixing intervals with very different signal-to-noise ratios (S/N), we first partitioned the observation into contiguous segments selected by the multi-harmonic H-test \citep{deJager:1989aa, dejager:2010aa}. The test was evaluated on unbinned photon phases up to $m_{\max}=10$ harmonics, and we retained only intervals with single-trial significance exceeding $5\sigma$. This procedure yielded 73 segments covering $\sim33$~ks (99.6\% of the good exposure), with a median duration of 445 s and a median of $\sim1065$ photons per segment. The median H-statistic is 36.3. The harmonic order that maximises the H-test varies mildly across the observation—most segments are consistent with a fundamental-only signal or a modest contribution from the second harmonic—indicating limited pulse-shape evolution and a profile dominated by the fundamental. A per-harmonic assessment with the unbinned $Z_k^2$ statistic (2 degrees of freedom; d.o.f. in the following) further shows that neither the second nor higher harmonics are individually significant at $\ge 3\sigma$ in any segment, so the segment-level signal is overwhelmingly carried by the fundamental.

For each valid segment, the pulse phase was obtained by maximising the unbinned periodicity statistic with respect to a simple phase shift, adopting the same number of harmonics, $m_\star$, that maximised the H-test \citep{Buccheri:1993aa}. The phase was measured relative to a high-S/N template built from the entire dataset and expressed as a truncated Fourier series with coefficients $T_k$ ($k \le m_\star$). For every segment, we calculated the complex Fourier coefficients $a_k$ from the photon phases and determined the shift that maximised their correlation with the template. This procedure is mathematically equivalent to an unbinned cross-correlation in phase space applied simultaneously to all harmonics. The statistical uncertainty for each phase was derived from non-parametric bootstrap resampling ($N_{\rm boot}=10^3$) of the photon list, preserving the signal's periodicity. The $1\sigma$ error was defined as the circular standard deviation of the recovered phase distribution. For a subset of segments, we also performed a parametric bootstrap based on simulated photon lists drawn from the Fourier template; the resulting scatter agreed within a few per cent with the non-parametric errors, supporting the reliability of the adopted estimates.

The resulting set of pulse phases was fitted with a small-eccentricity timing model using standard phase-coherent techniques \citep[see, e.g.,][]{Burderi:2007tl, Sanna:2016ty}. We started from the folding solution ($T_{\rm asc}=60805.36786(1)$~MJD, and $\nu=313.64373842(35)$~Hz), allowing differential corrections to all model parameters. Folding, phase extraction and the weighted fit were iterated until parameter updates were negligible compared to their formal uncertainties. Best-fit parameters from the coherent analysis are reported in Table \ref{table:timing}. As customary for AMXPs, we adopted parameter errors inflated by $\sqrt{\chi^2_{\rm red}}$ to account for any residual phase noise not captured by the statistical uncertainties \citep[][]{Finger:1999vb}. We also added, in quadrature, the systematic contribution to the spin frequency uncertainty due to the source position, following Eq.~(4) of \citet{Papitto:2011uv}. For our astrometry this term is $\simeq 3.2\times10^{-8}$~Hz and increases the total $1\sigma$ error by $<1\%$.

\begin{figure}
\centering
\includegraphics[width=0.48\textwidth]{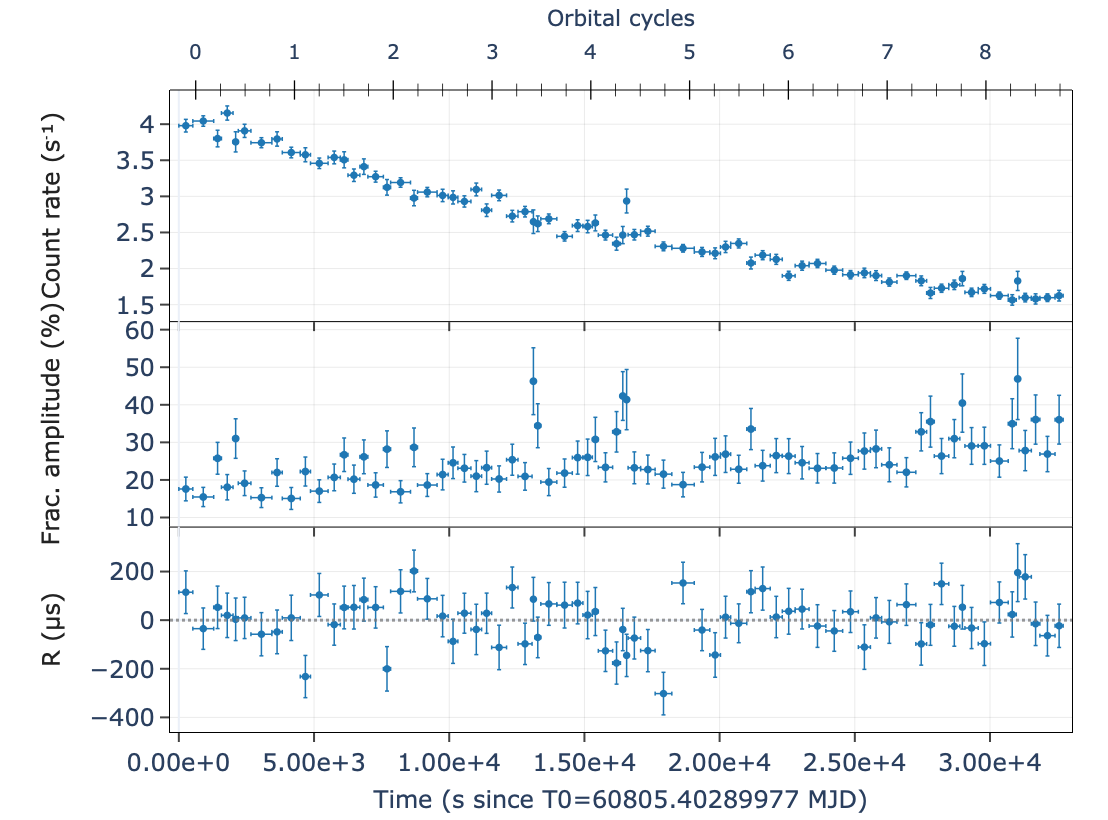}
\caption{Summary of the timing analysis for \maxi{} from the \xmm{}/EPIC-pn observation.
\textit{Top}: 0.5--10.0~keV background-subtracted count rate versus time for each time interval adopted to estimate a significant pulse profile.
\textit{Middle}: fractional amplitude of the fundamental harmonic in each H-test-selected segment; $1\sigma$ errors from photon bootstrap.
\textit{Bottom}: pulse phase residuals relative to the best-fitting coherent timing model (Table~\ref{table:timing}); the grey dashed line marks zero residual.
Time is measured from $T_{0}=60805.40289977$~MJD (TDB), \rev{and the horizontal axis along the top shows the corresponding orbital phase over the observation}.}
\label{fig:timing}
\end{figure}

The post-fit phase residuals for each H-test-selected segment show no systematic long-term drift across the observation (Fig.~\ref{fig:timing}, bottom panel). Their scatter is statistically consistent with the weighted coherent fit (Table~\ref{table:timing}), and augmenting the model with either a spin-frequency derivative $\dot{\nu}$ or an orbital-period derivative $\dot{P}_{\mathrm{orb}}$ does not yield a significant improvement. A modest excursion of the residuals is visible near the middle of the exposure; however, the phase promptly returns to the pre-excursion level, and the data require no persistent step. On the other hand, during the 2022 \nicer{} campaign \citep[see, Fig.~1 in][]{Sanna:2022vi}, a discrete phase jump was observed around MJD 59750.2, which persisted thereafter. Over the $\sim$33~ks timescale covered by the \xmm{}/EPIC-pn observation, the timing behaviour is therefore consistent with a circular orbit and a constant spin frequency within our sensitivity.

To quantify the pulse strength in each interval, we phase-aligned the events to the coherent solution. We fitted the mean-normalised binned profile with a harmonic model at the segment's optimal order $m_\star$. We take the fundamental amplitude to be the fractional-interval amplitude and estimate its $1\sigma$ uncertainty via a photon bootstrap. The fractional amplitude increases gradually as the source flux declines (Fig.~\ref{fig:timing}, middle panel), in agreement with the trend reported during the previous outburst \citep{Sanna:2022vi} and commonly seen in AMXPs \citep[e.g.,][]{Bult:2019tr, Bult:2022vn, Sanna:2022tt, Illiano:2023aa, Ballocco_2025arXiv}. Higher harmonics are not individually significant at $\geq 3\sigma$ in the segment-wise $Z_k^2$ analysis, indicating that, on short timescales, the observed variability reflects an overall rescaling of a profile dominated by the fundamental rather than a substantial change in pulse shape.

\begin{table}
\caption{\label{table:timing}Timing solutions for \maxi{} during the observed outbursts.}
\centering
\begin{tabular}{l c}
\hline
 & 2025 \\
\hline
Parameters  & \xmm{}\\
\hline
\hline
R.A. (J2000) & 19$^h$56$^m$39.11$^s$~$\pm$~0.02$^s$\\
DEC. (J2000) & 03$^\circ$26'43.7''~$\pm$~0.28''\\
$P_\mathrm{orb}$ (s) & 3653.21(24) \\
$x$ (lt-s) & 0.013828(18)\\
$T_{ASC}$ (MJD/TDB) & 60805.367834(16)\\
Eccentricity &  $<0.008$~(3$\sigma$~c.l.)\\
$T_0$ (MJD/TDB) & 60805.4\\
\hline
\hline
$\nu_0$ (Hz) & 313.64373844(34) \\
\hline
$\chi^2_\mathrm{red}$/dof & 1.33/66\\
\hline
\end{tabular}
\tablefoot{Orbital parameters and spin frequency evolution of \maxi{} obtained from the analysis of the \xmm{} observation from its latest outburst. $T_0$ represents the reference epoch for this timing solution. Uncertainties reported on the last digit correspond to the 1$\sigma$ confidence level. Errors are $1\sigma$ and scaled by $\sqrt{\chi^2_{\rm red}}$; the positional systematic on $\nu$ is added in quadrature and contributes $<1\%$. Source position from \citet{Chakrabarty:2016vp}}
\end{table}

Using the best-fitting ephemeris, we corrected the photon times and produced the average pulse profile. To maximise S/N, we scanned energy selections using an H-test-driven segmentation and adopted the $0.50-6.5$~keV band as the optimal range. The mean-normalised profile is smooth and largely dominated by the fundamental harmonic (H1). An unbinned H-test returns $m_{\rm opt}=5$, but only the first, second (H2),
and fifth (H5) harmonics are individually significant, while the remaining components are consistent with zero. A weighted least-squares refit to the binned profile reproduces the morphology well,
with the second harmonic accounting for the mild asymmetry of the waveform.

To characterise the energy dependence, we divided the spectrum into 16 significance-optimised
bands between $0.5$ and $10.0$~keV (each exceeding a fixed $\sigma$ selection threshold). In all bands, the signal is adequately described by a pure sinusoidal function ($m_\star=1$). From the best-fitting model, we measured the corresponding phase and fractional amplitude values. Figure~\ref{fig:energy_dep_phase} shows for the pulse phase a significant soft-lag trend with energy: a linear model improves over a constant by $\Delta \chi^2 = 10.12$ for one d.o.f. ($p=1.5\times10^{-3}$), with slope $b=-0.0123\pm0.0039$ cycles keV$^{-1}$ (relative to the reference band), and hints at a turnover above $\sim$3-4~keV where the errors increase. Figure~\ref{fig:energy_dep_amp} reveals that the fractional amplitude rises from $\sim$18-25\% below 1~keV to $\sim$30-33\% at 1.3-2.0~keV, then decreases above $\sim$3~keV to $\sim$16\% at 4.5-7.5~keV. A linear trend does not provide a significant improvement over a constant ($\Delta \chi^2 = 1.71$ for one d.o.f., null-hypothesis probability $p=0.19$), consistent with the non-monotonic shape.

\begin{figure}
\centering
\includegraphics[width=0.9\linewidth]{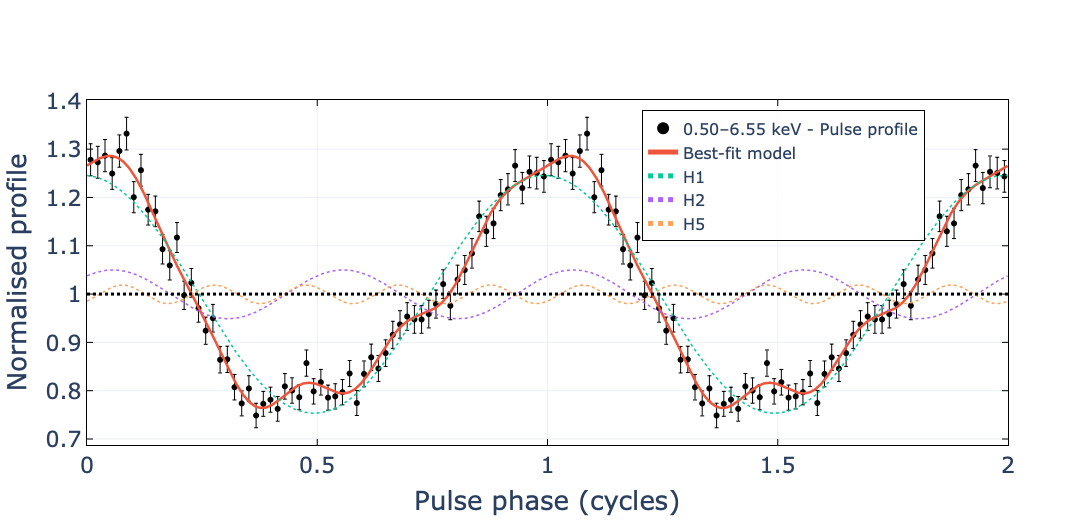}
\caption{
Pulse profile of \maxi{} in the 0.50--6.5~keV band.
Points (with $1\sigma$ errors) show the mean-normalised, 64-bin profile plotted over two cycles.
The solid line is the weighted least-squares fit at the H-test-selected order ($m_{\rm opt}=5$);
dotted curves display the individually significant harmonic components (H1, H2, H5).}
\label{fig:best_profile}
\end{figure}

\begin{figure}[t]
\centering

\subfloat[Evolution of the pulse phase delays as a function of energy.\label{fig:energy_dep_phase}]{
  \includegraphics[width=0.9\hsize]{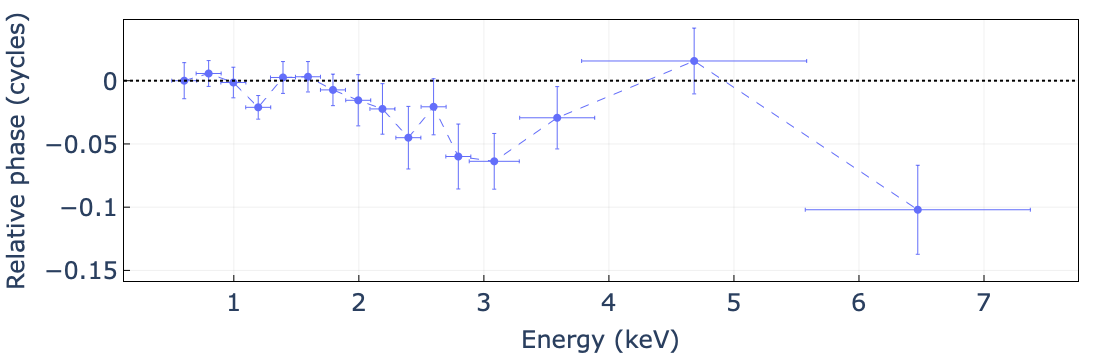}
}\\[2mm] 

\subfloat[Evolution of the pulse fractional amplitude as a function of energy.\label{fig:energy_dep_amp}]{
  \includegraphics[width=0.9\hsize]{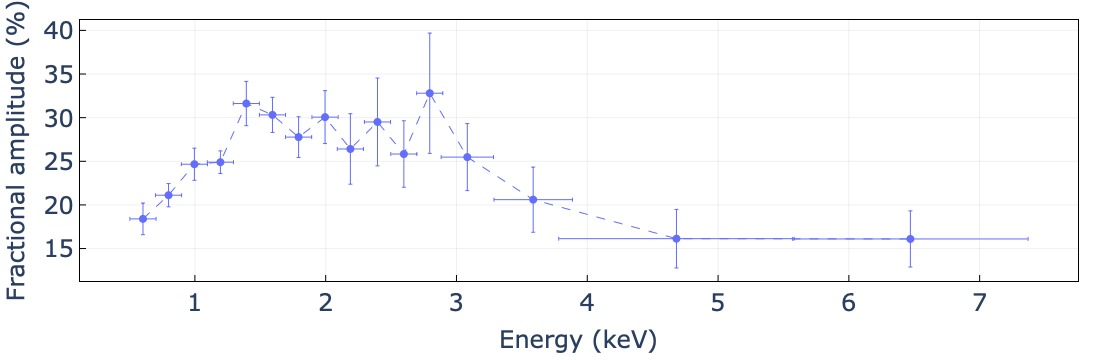}
}

\caption{
Energy dependence of the pulsations in \maxi{}. The \xmm{}/EPIC-pn dataset is split into 16 significance-optimised bands spanning 0.5--10.0~keV.
In each band, the signal is modelled with a pure sinusoidal function (fundamental harmonic) at the NS spin period.
Panel~(a): relative phase of the fundamental as a function of energy.
Panel~(b): fractional amplitude of the fundamental as a function of energy.
}
\label{fig:energy_dep}
\end{figure}


\subsection{Spectral analysis}
We performed the X-ray spectral analysis of \xmm{} and \swift{} data using the X-ray spectral fitting package \texttt{XSPEC} \citep{Arnaud96} version 12.14.1. We adopted the interstellar medium abundance and the cross-section tables from \citet{Wilms00} and \citet{Verner96}, respectively. All uncertainties on spectral parameters are given at the 1$\sigma$ confidence level.

\subsubsection{\xmm{} spectroscopy} \label{sec:XMM_spectrum}
We extracted the PN spectrum with a minimum of 25 counts in each channel. We limited the spectral analysis to the 0.5$-$8~keV band, as the background dominates outside this interval. 

We first fit the spectrum with an absorbed power-law model (\texttt{TBabs * powerlaw}), which yielded a $\chi^2$ = 213.27 for 128 d.o.f.. However, since AMXPs in outburst are typically described by thermally Comptonized emission \citep[e.g.,][]{DiSalvo_2023hxga.book}, we replaced the power-law component with the convolution model \texttt{thcomp} \citep{Zdziarski_2020MNRAS} applied to a blackbody component (\texttt{TBabs * (thcomp * bbodyrad)}; see red spectrum in Fig.~\ref{fig:spectra_comparison}). 
Given the lack of spectral coverage above 10~keV, we fixed the electron temperature, $\mathrm{kT_e}$, to 30~keV, consistent with the range of values found during the 2022 outburst by \citet{Sanna:2022vi}.
\rev{This revised model significantly improved the fit, yielding a final 
$\chi^{2}=141.24$ for 126 degrees of freedom (an improvement of $\Delta\chi^{2}\simeq72$ with two fewer free parameters compared to the power-law model). The resulting best-fit parameters are listed in Table~\ref{tab:params_spectrum}. The unabsorbed 0.5$-$10~keV flux, $F_\mathrm{{0.5-10}}$, was estimated by including the \texttt{cflux} component. In the context of the full outburst evolution, the EPIC-pn spectrum is consistent with the softening trend seen in the \swiftxrt{} data (Sect.~\ref{sec:Swift_spectroscopy_analysis}), showing a relatively steep Comptonised continuum and a comparatively cool blackbody component, as commonly observed in AMXPs during outburst decay.} 

Given that the reflection spectrum, especially the iron K$\alpha$ complex, is commonly observed in AMXPs in outburst (see, e.g., Table 3 from \citealt{Illiano:2024aa} and references therein), we searched for similar features in \maxi{} by adding Gaussian lines to our best-fit model. We fixed the other parameters to their previously obtained values.
The line energies were set consecutively at 6.4~keV (neutral or lightly ionised Fe~K$\alpha$), 6.7~keV (He-like Fe~XXV), and 6.97~keV (H-like Fe~XXVI).
To account for possible reflection broadening, we adopted a Gaussian width of $\sigma = 0.2$~keV, slightly larger than the intrinsic spectral resolution of EPIC-pn, and typical of Fe lines observed in reflection spectra of similar systems (e.g., \citealt{Illiano:2024aa} and references therein). No significant emission feature was found. We derived 3$\sigma$ upper limits on the equivalent widths of each of these features at 6.4, 6.7, and 6.97~keV, which were $\sim$0.25, 0.29, 0.23~keV, respectively.

\begin{table} 
\renewcommand{\arraystretch}{1}
\centering
\caption{Best-fit spectral parameters from the model \texttt{TBabs * (thComp*bbodyrad)} for the \xmm{}/EPIC-pn spectrum.} \label{tab:params_spectrum}
\begin{tabular}{l c c}          
   \hline\hline
    Component & Parameter & Value \\
    \hline
    \scshape{Tbabs} & $\mathrm{N_H}$ ($10^{21}$ cm$^{-2}$) & $0.3 \pm 0.1$\\
    \scshape{thComp} & $\Gamma$ & $2.4 \pm 0.1$ \\
    & $\mathrm{kT_e}$ (keV) & $30^{(*)}$ \\
    & $\mathrm{cov\_frac}$ & $0.54 \pm 0.06$\\
    \scshape{bbodyrad} & $\mathrm{kT}$ (keV) & $0.226 \pm 0.007$ \\
    & $\mathrm{Norm_{bbodyrad}}$ & $86^{+11}_{-9}$\\
    & $\mathrm{R_{bbodyrad}}$ (km) & $4.6^{+0.7}_{-0.6}$\\
    \scshape{cflux} &$F_\mathrm{{0.5-10}} \, \mathrm{(10^{-12} \, erg \, cm^{-2} \, s^{-1})}$ & $5.72 \pm 0.08$\\
    &$F_\mathrm{{0.5-2}} \, \mathrm{(10^{-12} \, erg \, cm^{-2} \, s^{-1})}$ & $3.70 \pm 0.08$\\
    &$F_\mathrm{{2-10}} \, \mathrm{(10^{-12} \, erg \, cm^{-2} \, s^{-1})}$ & $2.03 \pm 0.04$\\
    \hline
    &$\chi ^2$/d.o.f & 141.24/126\\
    \hline
\end{tabular}
\tablefoot{$N_\mathrm{H}$ is the absorption column density, $\Gamma$ is the photon index, $\mathrm{kT_e}$ is the electron temperature, $\mathrm{cov\_frac}$ is the covering fraction, and $\mathrm{Norm_{bbodyrad}}$ is the normalization for the \texttt{bbodyrad} component.
The radius of the emitting region, $\mathrm{R_{bbodyrad}}$, in km was estimated using the relation $\mathrm{Norm_{bbodyrad}} = \mathrm{R_{bbodyrad}}^2 d_{10}^2$, where $d_{10}$ is the distance to the source in units of 10 kpc. We adopted $d = 5 \pm 2$ kpc from \citet{Ravi:2017tl}.
$F_\mathrm{{0.5-10}}$, $F_\mathrm{{0.5-2}}$, and $F_\mathrm{{2-10}}$ are the unabsorbed fluxes estimated in the 0.5$-$10~keV, in the 0.5$-$2~keV, and in the 2$-$10~keV energy bands, respectively. All uncertainties are given at the 1$\sigma$ confidence level.
$^{(*)}$ Kept frozen during the fit.}
\end{table}
\noindent

\subsubsection{\swift{} spectroscopy}  \label{sec:Swift_spectroscopy_analysis}
\begin{figure}
    \centering
    \includegraphics[width=0.4\textwidth]{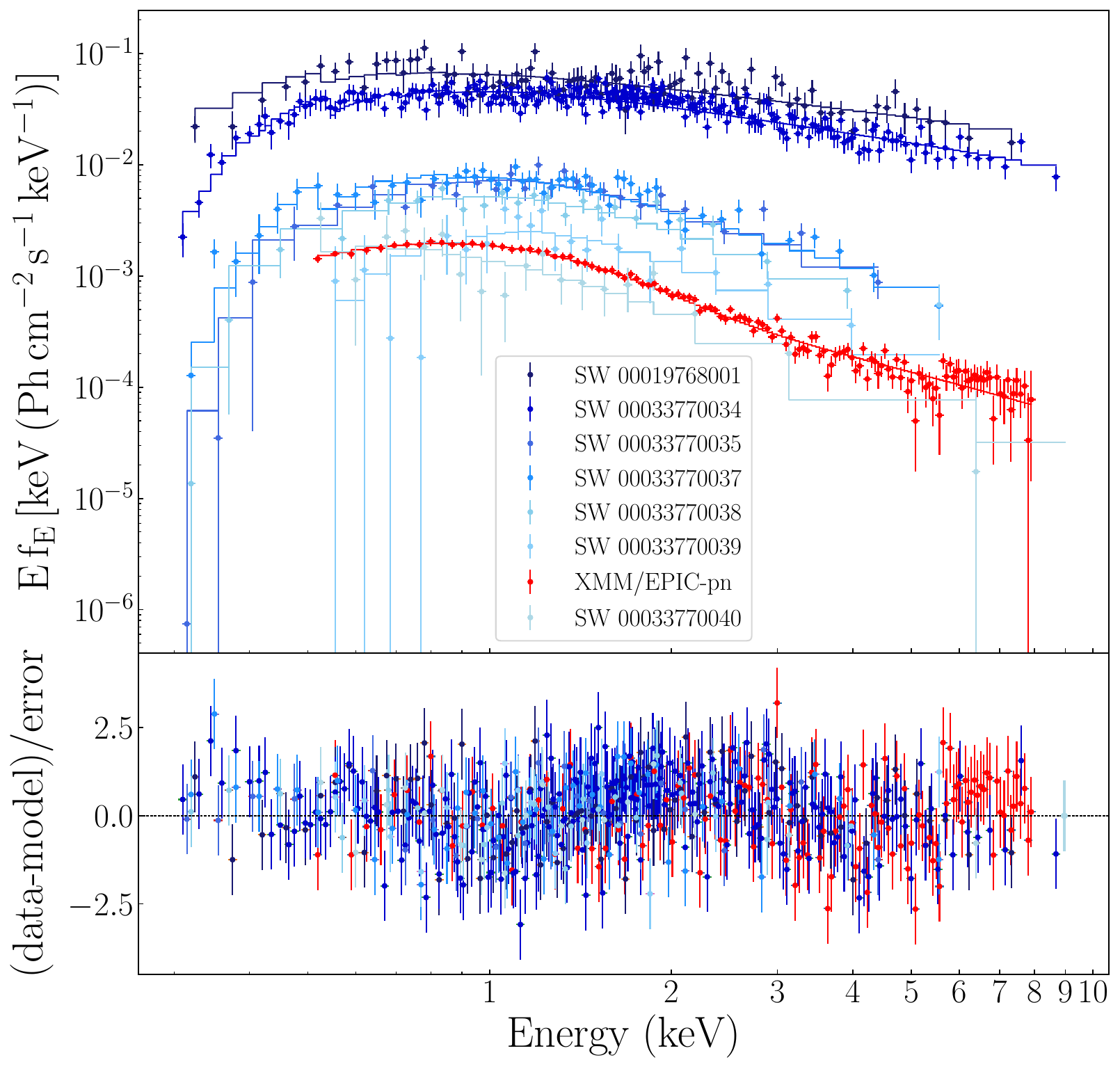}
    \caption{Unfolded average spectrum of the persistent emission from \maxi. The blue curves represent \swift{}/XRT spectra, from the earliest observation (darkest blue) to the latest (lightest blue), fitted with the \texttt{TBabs*powerlaw} model (see Sect.~\ref{sec:Swift_spectroscopy_analysis}). In the legend, \swift{}/XRT observations are labelled as ``SW" followed by their ObsID (see Table~\ref{tab:swift_obs}). The red curve shows the \xmm{}/EPIC-pn spectrum, fitted with \texttt{TBabs*(thComp*bbodyrad)} (see Sect.~\ref{sec:XMM_spectrum}). The bottom panel displays the residuals with respect to the adopted models.} \label{fig:spectra_comparison}
\end{figure}

\begin{figure}
    \centering
    \includegraphics[width=0.35\textwidth]{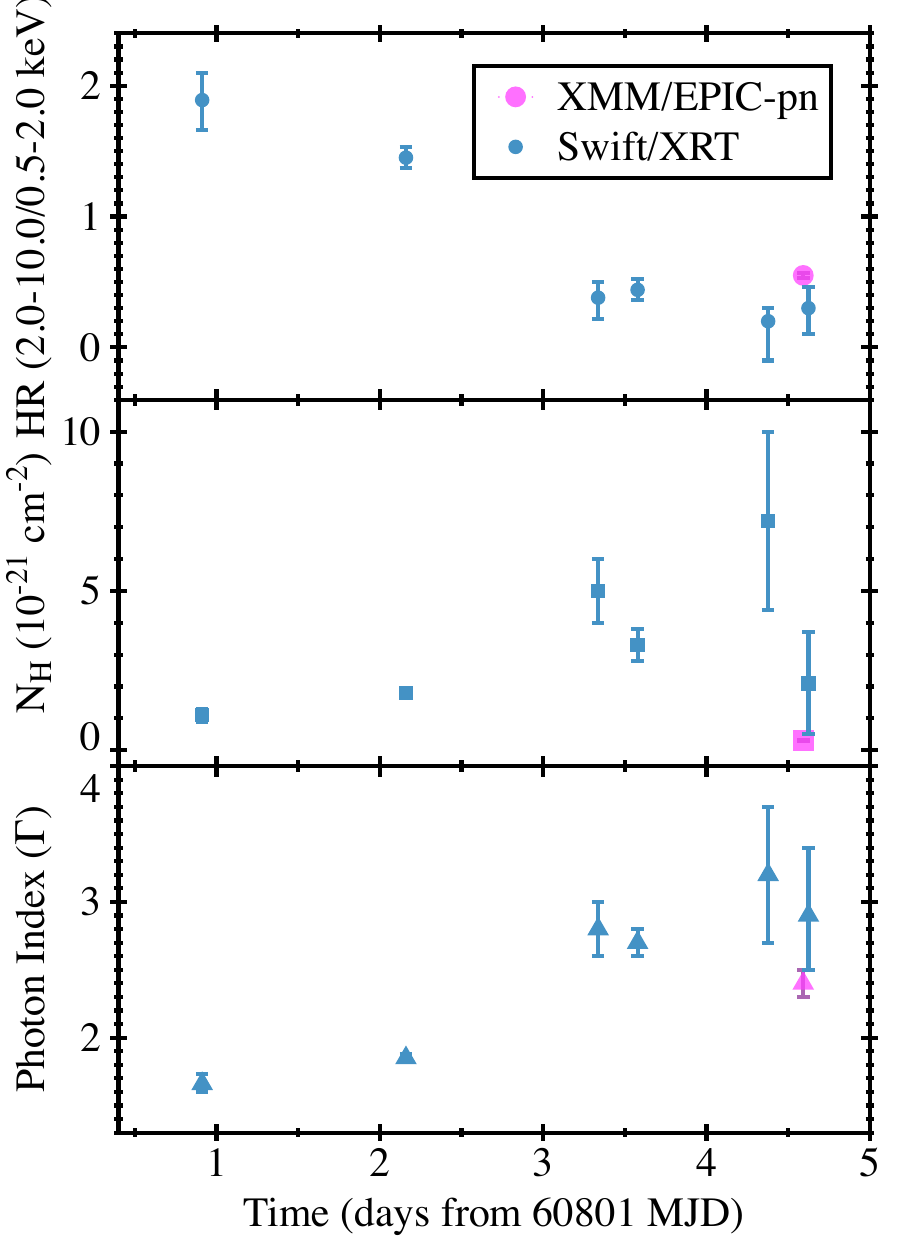}
    \caption{\rev{Temporal evolution of the main spectral parameters describing the continuum emission of \maxi{} observed with \swift{}/XRT (blue points) and \xmm{}/EPIC-pn (pink points). The top panel shows the hardness ratio $\mathrm{HR}=F_\mathrm{{2-10}}/F_\mathrm{{0.5-2}}$, the middle panel shows $N_\mathrm{H}$, and the bottom panel displays the photon index $\Gamma$. All values are listed in Tables~\ref{Table:spectra_Swift} and \ref{tab:params_spectrum}. Error bars represent 1$\sigma$ uncertainties.}} \label{Fig:spectral_evolution}
\end{figure}

We modelled the first \swift{}/XRT spectrum acquired in PC mode at the onset of the outburst (see Table~\ref{tab:swift_obs}). An absorbed power-law model (\texttt{TBabs * powerlaw}) provided a good fit with a $\chi^2$ of 93.23 for 93 d.o.f.. Motivated by the results obtained for the \xmm{}/EPIC-pn spectrum (Sect.~\ref{sec:XMM_spectrum}), we also tested the \texttt{TBabs * (thcomp * bbodyrad)} model, fixing $\mathrm{kT_e}$ to 30~keV and the $\mathrm{N_H}$ to the value derived from the power-law model, since the other parameters were otherwise unconstrained. However, the fit did not provide meaningful constraints, with a blackbody normalisation consistent with zero within 2$\sigma$. For this reason, we adopted the simpler power-law model to describe all XRT spectra in order to track the spectral evolution throughout the outburst consistently (see blue spectra in Fig.~\ref{fig:spectra_comparison}). 
The best-fit parameters are reported in Table~\ref{Table:spectra_Swift} and shown in Fig.~\ref{Fig:spectral_evolution}. All spectral parameters are consistent within 1$\sigma$ with those obtained with the online XRT product generator, except the $\mathrm{N_H}$ value of ObsID 00033770035, which is consistent within 3$\sigma$.
The last \swift{}/XRT observation analysed (ObsID 00033770040), which is the one closest in time to the \xmm{} observation, shows a slight discrepancy in the spectral parameters compared to the PN spectrum (Fig.~\ref{Fig:spectral_evolution}). However, both the $\mathrm{N_H}$ and the $\Gamma$ are consistent within 1.5$\sigma$, also due to the large uncertainties associated with the low statistics of the \swift{} spectrum.

By including the \texttt{cflux} component, we estimated the X-ray unabsorbed flux in three bands: 0.5$-$10~keV flux ($F_\mathrm{{0.5-10}}$), 0.5$-$2~keV flux ($F_\mathrm{{0.5-2}}$; soft band), 2$-$10~keV flux ($F_\mathrm{{2-10}}$; hard band). \rev{From these quantities we computed, for each observation, the hardness ratio $\mathrm{HR}=F_\mathrm{{2-10}}/F_\mathrm{{0.5-2}}$; the temporal evolution of HR, together with that of $N_\mathrm{H}$ and $\Gamma$, is shown in Fig.~\ref{Fig:spectral_evolution}. The hardness ratio shows a clear softening trend as the outburst decays, decreasing steadily at lower fluxes. This behaviour closely resembles the spectral evolution seen during the 2022 outburst and indicates a smooth spectral softening during the decay rather than a full hard-to-soft state transition.} 

\subsection{Optical photometry}
\maxi{} was clearly detected in all unblended LCO images obtained between 2025 May 8 and 11 (MJD 60803–60806).
The calibrated magnitudes used here are listed in Appendix \ref{app:optical}.
\rev{A joint X-ray/optical view of the 2025 outburst is shown in Fig.~\ref{fig:xray_opt_lc}, where the \swift{}/XRT 0.5--10~keV light curve is plotted together with the LCO $g'$, $r'$, and $i'$ photometry.}
The source reached its maximum optical brightness on 2025 May 10 (MJD 60805; see also \citealt{Illiano_2025ATel17187}), after which it declined rapidly.
Subsequent LCO observations on May 13 (MJD 60808.03) showed a blended magnitude of $g’ = 20.30 \pm 0.20$, consistent with the nearby field star and indicating that the outburst had ended at optical wavelengths.

At its peak, the 2025 outburst did not reach the same brightness level as the first peak of the previous outburst on 2022 June 20 (MJD~59750) \citep[][; see also Fig. \ref{fig:2022outburst}]{Atel15448,Baglio2022_2}, when \maxi{} reached $g' = 19.087 \pm 0.016$~mag, $i' = 18.943 \pm 0.034$~mag (the peak in $r'$ band was instead reached on MJD ~59751, with $r' = 18.923\pm 0.086$~mag).
The optical evolution during the 2025 outburst does not appear to track the X-ray flux.
While the optical flux increased slightly between MJD~60803 and MJD~60805, the 2–10~keV X-ray flux declined by more than an order of magnitude (a factor of $\sim$14; Table~\ref{tab:params_spectrum}).
This apparent anti-correlation suggests that X-ray reprocessing is unlikely to dominate the optical emission during this phase.
Instead, the optical light may originate primarily from the intrinsic thermal emission of the outer accretion disc.
Although no optical data are available before MJD~60803, it is possible that an earlier optical maximum occurred prior to our observing window.
Indeed, \citet{Kong_2025ATel17171} reported optical brightening of the source as early as 2025~May~5 (MJD~60800), preceding the first X-ray observations, although no magnitudes were provided.
Taken together, these results indicate that the optical emission evolved independently of the X-rays, with the optical maximum likely occurring after substantial X-ray fading.

\begin{figure}
    \centering
    \rev{\includegraphics[width=0.8\linewidth]{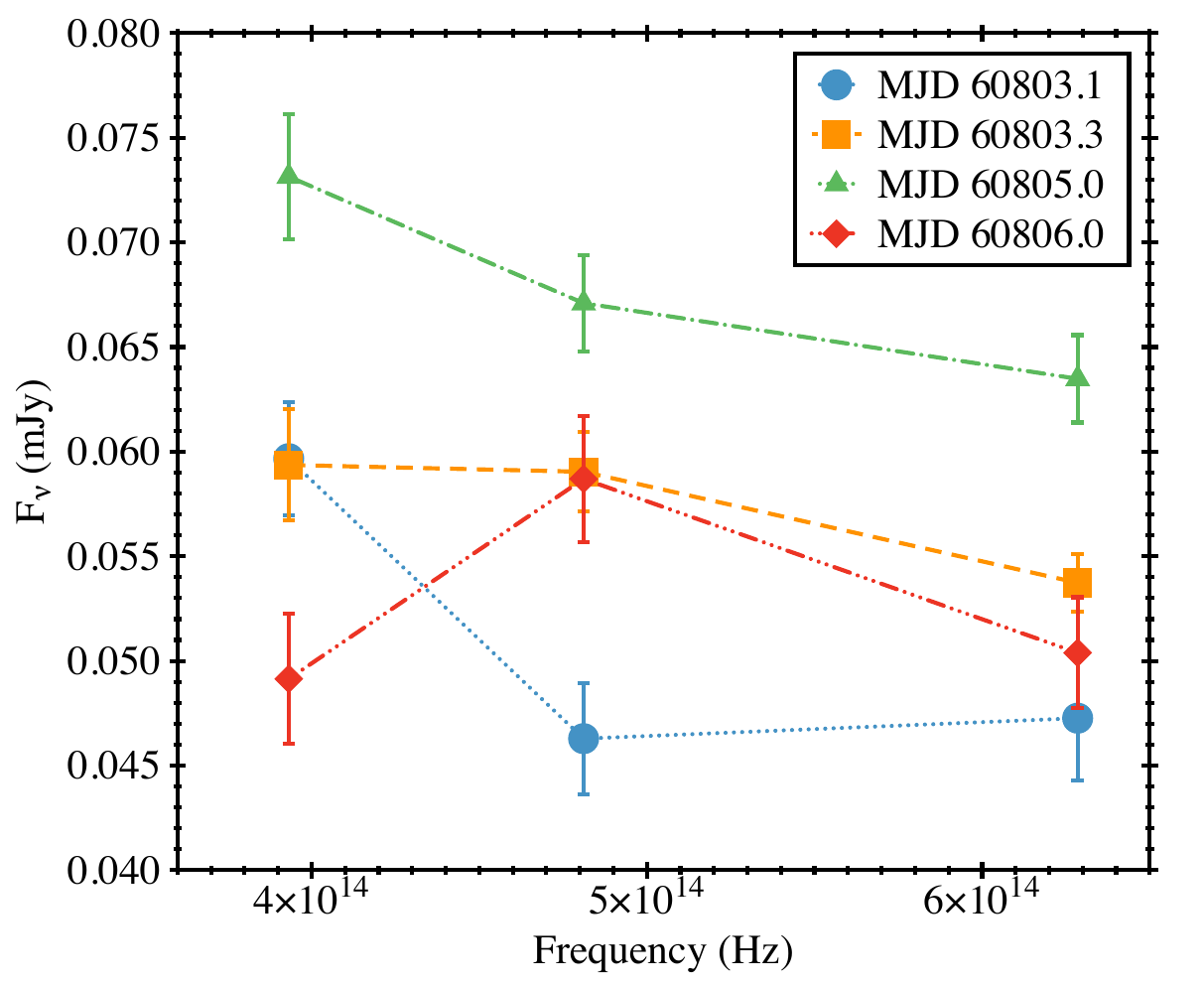}}
    \caption{Spectral energy distributions of \maxi{} obtained from LCO observations between 2025~May~8 (MJD 60803) and 2025~May~11 (MJD 60806). Different colours and symbols correspond to the observation dates (MJDs 60803.1, 60803.3, 60805.0, and 60806.0), and the error bars indicate the propagated uncertainties, including both photometric and reddening contributions.}
    \label{fig:SEDs_opt}
\end{figure}

Figure~\ref{fig:SEDs_opt} shows the nearly simultaneous optical spectral energy distributions (SEDs) of \maxi{} obtained between 2025~May~8 and 2025~May~11.
The flux densities have been corrected for Galactic reddening using the \texttt{dust\_extinction} Python package with extinction curves from \citet{Gordon2024} \citep[see also][]{Gordon_2009, Fitzpatrick_2019, Gordon_2021, Decleir_2022}, adopting a colour excess of $E(B-V)=0.034\pm0.011$~mag derived from the neutral hydrogen column density $N_{H}=(0.3\pm0.1)\times10^{21}\, \rm cm^{-2}$ and using the \cite{Foight2016} conversion (Table~\ref{tab:params_spectrum}). This value corresponds to the lowest $N_{H}$ measured from the \xmm{} spectrum, which is likely to represent the foreground absorption best; intrinsic variations in $N_{H}$ seen during the outburst are not expected to correlate with optical extinction \citep[e.g.,][]{Oates2019}.
The SEDs indicate that the outburst reached its optical peak between MJD~60803.3 and MJD~60805, followed by a rapid decline thereafter.
A clear colour evolution is also evident, with the $i'$-band flux decreasing more strongly between MJD~60805 and MJD~60806 than in the bluer bands.

\section{Discussion and Conclusions}

We presented the temporal and spectral properties of the
AMXP \maxi{} during its 2025 outburst by analysing the available \xmm{}, \swift{} and optical datasets. 

\subsection{Timing analysis}
We performed a phase-coherent timing analysis of \maxi{} during its 2025 flash outburst, showing that the source retained a remarkably stable rotational ephemeris over the \xmm{} exposure. Within sensitivity, no spin-frequency derivative or orbital evolution is required, and the post-fit residuals display no persistent steps (Fig.~\ref{fig:timing}). This behaviour contrasts with the 2022 \nicer{} campaign, where a discrete, long-lived phase discontinuity appeared near MJD~59750.2 \citep{Sanna:2022vi}, and no analogous feature is required here. This suggests that phase irregularities in \maxi{} are sporadic manifestations of disc–magnetosphere instabilities that reconfigure the magnetospheric coupling and shift the accretion footprint, producing phase offsets and rapid profile changes \citep[see, e.g.,][]{Riggio:2008wz, Patruno:2009vg, Patruno2009a, Kajava2011a, Poutanen:2009wb, Ibragimov2009a}. A similar intermittency of phase irregularities has been reported in the prototypical AMXP SAX~J1808.4$-$3658, where profile evolution and occasional phase jumps complicate torque measurements \citep[see, e.g.,][]{Burderi:2006va, Hartman:2008uj}. In the broader AMXP population, low-level timing noise often tracks the X-ray flux via phase–flux correlations, suggesting a connection with a hotspot in motion and a corresponding bias on apparent spin-frequency derivatives $\dot{\nu}$ \citep[see, e.g.,][]{Patruno:2009vg}. The absence of a measurable $\dot{\nu}$ here is therefore unsurprising, given both the short baseline and the near-constancy of the phase residuals. Single-outburst torque estimates based on short-baseline timing should, in general, be interpreted with caution \citep[see, e.g.,][]{Hartman:2008uj, Patruno:2009vg}.

\subsubsection{Pulse profile and energy dependence}

On short timescales, the \maxi{} profiles are driven almost entirely by the fundamental component, and the segment-to-segment variability is well described by amplitude rescaling rather than genuine shape changes. The fractional amplitude gradually increases as the flux declines, mirroring the previous outburst and trends seen in other AMXPs, and suggesting that the emitting region becomes more localised as the accretion rate drops \citep[see, e.g., MAXI J1816-195, IGR J17379-3747, IGR J17498-2921][]{Bult:2019tr, Bult:2022vn, Illiano:2024aa}. For \maxi{}, the combination of (i) an essentially sinusoidal waveform, (ii) an amplitude–flux anti-correlation, and (iii) the reproducibility of the shape between 2022 and 2025 points to a stable visibility geometry for the hotspot and a relatively quiescent disc–magnetosphere coupling over the \xmm{} interval.

By combining the whole dataset, we obtained the most significant average pulse profile in the optimal 0.50–6.5~keV band. The morphology reinforces this view of geometrical steadiness. A noteworthy feature of the average profile is the significant detection of the fifth harmonic, while the third and fourth remain undetected within tight upper limits. Rich harmonic content is not unprecedented among AMXPs, but this specific configuration is. In several systems higher-order harmonics have been detected, though usually with rapidly decreasing amplitudes: in IGR~J17591$-$2342 the \nicer{} profile required four components, but in energy-resolved profiles, only the fundamental and second harmonic remained significant \citep{Sanna:2020wv}; SWIFT~J1749.4$-$2807 showed harmonics up to third order, with the second harmonic even exceeding the fundamental, consistent with two nearly antipodal spots at high inclination \citep{Sanna:2022tt}; the ultra-compact IGR~J16597$-$3704 also exhibited up to four harmonics in \nustar{} data \citep{Sanna:2018td}. These examples show that AMXP waveforms can carry power to high orders, but detectability is strongly controlled by geometry, anisotropy and S/N, providing a natural context for our case, where the fifth harmonic is detected while the third and fourth are not.

From a theoretical standpoint, \citet{Poutanen:2006aa} showed that the Fourier content of the waveform is governed by (i) the number and geometry of emitting regions (primary and antipodal spots), (ii) beaming anisotropy (e.g. a Comptonising slab), and (iii) relativistic effects (Doppler boosting, aberration and light bending). In their analytic hotspot framework, specific harmonics can be strongly suppressed for plausible viewing geometries and anisotropy; in particular, anisotropy naturally enhances the second harmonic, while time-delay effects generate a small third harmonic. Complementary calculations by \citet{Viironen:2004aa} show how a Comptonising slab can imprint narrow phase substructure whose power is distributed into higher orders, with a mild energy dependence.

In this context, two scenarios could qualitatively reproduce our findings. In a \emph{two-spot interference} picture, two nearly antipodal spots with similar brightness partly cancel odd harmonics; small departures from perfect antipodality and modest asymmetries in intensity or longitude can shift the cancellation so that the third and fourth harmonics lie near minima, while a residual fifth harmonic remains above our sensitivity. Alternatively, an \emph{anisotropic beaming} scenario invokes a weak, phase-confined excess from a Comptonising layer (e.g. an accretion shock just above the hotspot) that seeds higher-order terms without lifting all orders equally \citep{Viironen:2004aa}. The resulting phase-localised anisotropy provides a natural mechanism to enhance specific higher-order harmonics while maintaining minima at intermediate orders \citep{Poutanen:2006aa}.

While these scenarios allow selective suppression of specific harmonics, the precise pattern observed here is uncommon for AMXPs, so our interpretation remains provisional. A more secure conclusion will require higher photon statistics and source-specific relativistic two-spot modelling (e.g. allowing small offsets in spot longitudes and brightness and an explicit anisotropy parameter), together with harmonic-by-harmonic tests (e.g. stability of the fifth-harmonic phase and tighter upper limits on the lower components).

The pulse shape of \maxi{} remains essentially sinusoidal at all energies, with variability dominated by the fundamental and only weak higher-order structure. The fractional amplitude increases as the outburst fades and, within each dataset, shows the familiar mild rise with photon energy before flattening, a behaviour seen in several AMXPs and usually attributed to the changing mix of a hard Comptonised component and a softer thermal contribution from the hotspot and/or disc. A similar amplitude–energy trend (rise to a few keV, then plateau or mild decline) together with soft lags that grow with energy has been observed in other AMXPs. \nicer{} observations of IGR~J17591$-$2342, for example, showed a pattern well reproduced by down-scattering of hard photons in relatively calm plasma near the surface or disc \citep[see Fig.~4 and discussion in][]{Sanna:2020wv}. Our energy-resolved timing analysis is consistent with the standard hotspot plus Comptonisation picture developed for AMXPs: a hard, mildly anisotropic Comptonised beam leads in phase and sets the amplitude–energy rise, while a softer component lags due to reprocessing or scattering \citep[see, e.g.,][]{Cui98b, Poutanen:2003aa, Falanga:2007ub}. The close similarity between the 2025 behaviour and the 2022 outburst (excluding the phase-jump interval) argues for a stable viewing geometry and magnetospheric coupling across events.

\subsubsection{Long-term spin evolution}
\label{sec:spin}

By adopting the spin frequency obtained from the coherent 2022 timing solution reported by \citet{Sanna:2022vi} on Table 1 ($\nu_{22}=313.64374049(22)\ \mathrm{Hz}$ at $T_{0}=59749.0\ \mathrm{MJD\ (TDB)}$), and comparing it with the 2025 \xmm{} timing solution ($\nu_{25}=313.64373844(34)\ \mathrm{Hz}$ at $T_{0}=60805.4\ \mathrm{MJD\ (TDB)}$), we find a small but significant spin-down over the $\Delta t \simeq 1056~\mathrm{d}$ between the two epochs. The frequency offset is
\begin{equation}
\Delta\nu = \nu_{25} - \nu_{22} = (-2.05 \pm 0.04)\times10^{-6}\ \mathrm{Hz},    
\end{equation}
corresponding to an average spin derivative of
\begin{equation}
\langle \dot{\nu} \rangle = \frac{\Delta\nu}{\Delta t} = (-2.25 \pm 0.44)\times10^{-14}\ \mathrm{Hz\ s^{-1}}.
    \end{equation}
The magnitude of $\langle\dot\nu\rangle$ is about an order of magnitude larger (in absolute value) than the secular spin-down reported for sources such as SAX J1808.4−3658 \citep[$\dot\nu\simeq -1.152(56)\times10^{-15}\ \mathrm{Hz\ s^{-1}}$][]{Illiano:2023aa}, IGR J00291+5934 \citep[$\dot\nu\simeq -4.1(1.2)\times10^{-15}\ \mathrm{Hz\ s^{-1}}$][]{Patruno:2010tm, Papitto:2011uv}, XTE J1751-305 \citep[$\dot\nu\simeq -5.5(1.2)\times10^{-15}\ \mathrm{Hz\ s^{-1}}$][]{Riggio:2011vv}, IGR
J17498-2921 \citep[$\dot\nu\simeq -4.1(2)\times10^{-15}\ \mathrm{Hz\ s^{-1}}$][]{Illiano:2024aa}, IGR J17511-3057 \citep[$\dot\nu\simeq -2.3(1.1)\times10^{-15}\ \mathrm{Hz\ s^{-1}}$][]{Sanna:2025ab}, but comparable to what has been reported for the AMXP IGR J17494-3030 \citep[$\dot\nu\simeq -2.1(7)\times10^{-14}\ \mathrm{Hz\ s^{-1}}$][]{Ng:2021aa}. 

The 2022 \nicer{} timing analysis revealed a $\sim0.2$~cycle phase jump during the outburst \citep{Sanna:2022vi}, which was modelled through a phase step in their coherent solution. The single-spin frequency reported in their Table 1 likely corresponds to the most stable portion of the signal after that discontinuity and is therefore adopted here as the reference value for the long-term comparison. Given the occurrence of phase jumps, the derived frequency and, hence, the inferred spin-down derivative should be treated with caution and verified through future monitoring across multiple outbursts.

Even with this limitation, the observed spin-frequency derivative offers a means to quantify the torque acting on the NS and to compare it with values measured in other AMXPs.
From the long-term spin derivative derived, we can estimate the braking torque on the star following from angular-momentum conservation
\begin{equation}
N \;=\; I\,\dot\Omega \;=\; 2\pi\,I\,\dot\nu,    
\end{equation}
where $I$ is the NS moment of inertia. Adopting a canonical $I=10^{45}\ {\rm g\,cm^{2}}$ and the observed $\langle\dot\nu\rangle\simeq-2.25\times10^{-14}\ {\rm Hz\,s^{-1}}$, we infer
\begin{equation}
N \;=\; 2\pi \times 10^{45}\times(-2.25\times10^{-14}) \;\simeq\; -1.4\times10^{32}\ {\rm dyne\,cm}.    
\end{equation}
If we interpret the secular $\dot\nu$ as dominated by magnetic-dipole braking during quiescence, the polar dipole field ($B_{\rm p}$) can be estimated from the free-force dipole torque \citep[orthogonal rotator][]{Spitkovsky:2006uz}
\begin{equation}
N_{\rm ff} \;=\; -\frac{\mu^{2}\,\Omega^{3}}{c^{3}(1+\sin^2 \alpha)},\qquad
\mu \equiv \frac{B_{\rm p}R^{3}}{2},\qquad \Omega=2\pi\nu,
\end{equation}
which gives the standard frequency-space relation
\begin{equation}
\dot\nu \;=\; -\,\frac{2\pi^{2}}{3\,I\,c^{3}}\, B_{\rm p}^{\,2}\,R^{6}\,\nu^{3}\,(1+\sin^{2}\!\alpha),
\end{equation}
hence
\begin{equation}
B_{\rm p}\;=\;\left[\frac{I\,c^{3}\,|\dot\nu|}{\pi^{2}R^{6}\nu^{3}\,(1+\sin^{2}\!\alpha)}\right]^{1/2}\;,    
\end{equation}
where $\alpha$ is the magnetic obliquity (angle between the rotation and magnetic axes). \rev{With $R\approx 10\ {\rm km}$ and $\alpha=90^\circ$, this gives a characteristic field of order \rev{$B_{\rm p}\lesssim 1\times10^{9}\ {\rm G}$}. We stress that this dipole estimate should be regarded as an upper limit}, since part of the measured secular spin-down may be produced by residual disc torques or timing-noise/phase–flux systematics rather than pure dipole braking \citep[see, e.g., phase-flux bias discussion][]{Patruno:2009vg}. The surface dipolar field inferred for \maxi{} is consistent with that reported for other AMXPs \citep{Mukherjee:2015td}. 

\subsection{Spectral properties}
AMXPs typically exhibit hard X-ray spectra during their outbursts, usually without showing transitions to softer states \citep[see, e.g.,][and references therein]{Poutanen_2006AdSpR, DiSalvo_2023hxga.book}. The spectral shape is often characterised by a power-law emission, with photon indices commonly in the range $\Gamma \sim 1.8-2.0$. This hard X-ray emission is usually attributed to thermal Comptonisation processes, where seed blackbody-like photons with a temperature of $\sim$0.3$-$1.0~keV are up-scattered by hot electrons at temperatures of $\sim$20$-$50~keV \citep[see, e.g.,][]{Papitto_2020NewAR}. Notably, AMXPs tend to display reduced spectral variability during outbursts when compared to non-pulsating NS in LMXBs \citep[e.g.,][]{Illiano:2024aa, Li_2024AA, manca:2023aa, Marino_2022MNRAS}.

In our analysis, the \xmm{}/EPIC-pn spectrum of \maxi{} was well fitted using a thermal Comptonisation model (\texttt{thcomp}) convolved with a blackbody component, in line with findings from the 2022 outburst \citep{Sanna:2022vi}. Adopting a distance of 5~kpc \citep{Ravi:2017tl}, we derived a blackbody normalisation corresponding to a radius of $4.6^{+0.7}_{-0.6}$ km, consistent with emission from a localised region such as a hotspot on the NS surface, as previously reported.

Unlike the majority of AMXPs, which typically exhibit persistently hard X-ray spectra throughout their outbursts as mentioned above, \maxi{} showed a marked spectral softening during its 2025 event, despite the outburst lasting only about five days. \rev{This was evident in the evolution of the HR, which dropped from $\sim$2 at the onset to $\sim$0.2 in the decay phase (see the top panel of Fig.~\ref{Fig:spectral_evolution}), and in the progressive increase of the photon index from $\sim$1.7 to values approaching 3 (bottom panel of Fig.~\ref{Fig:spectral_evolution} and Table~\ref{Table:spectra_Swift}).} This spectral softening closely mirrors the behaviour observed during the 2022 outburst, when \citet{Sanna:2022vi} reported a photon index reaching up to $\sim$2.8 during the decay phase \citep[see, also][]{beri:2019aa, Wijnands:2015aa, manca:2023aa}. \rev{Although \maxi{} reaches luminosities higher than those of canonical VFXTs, during most of the decay it traverses the $L_X\sim10^{34}$--$10^{36}$~erg~s$^{-1}$ range where NS VFXTs are known to soften. The increase of the photon index from $\sim$1.7 to $\sim$3 is fully compatible with the $L_X$-$\Gamma$ behaviour observed in these systems \citep[see, e.g.,][]{Wijnands:2015aa}.}

\rev{We also observed variability in the column density ($N_\mathrm{H}$) throughout the outburst (middle panel of Fig.~\ref{Fig:spectral_evolution}), consistent with the behaviour seen in previous episodes.} The $N_\mathrm{H}$ values obtained from the \textit{Swift}/XRT spectra are systematically higher than those derived from the \textit{XMM-Newton}/EPIC data, likely because the simple power-law model used for \textit{Swift} does not capture the intrinsic low-energy curvature accounted for by the \texttt{thcomp*bbodyrad} model adopted for \textit{XMM-Newton}.
Finally, no disk reflection features were detected in the spectra, again in line with the results reported for the 2022 outburst \citep{Sanna:2022vi}. Reflection signatures have been observed in many AMXPs \citep[see, e.g., Table 3 in][]{Illiano:2024aa}, however, several systems, like \maxi, show no clear evidence of such components \citep[e.g.,][]{Falanga_2005A&A, Sanna:2018td}. The lack of detectable Fe~K emission may be because the emitting region is small and illuminated by a concentrated X-ray flux from the hotspot or accretion column, leading to a very high ionisation that prevents the formation of discrete lines \citep[see, e.g.,][]{Ross05}.

Recently, coherent X-ray pulsations have been detected from the two AMXPs IGR J17511$-$3057 and IGR J17379$-$3747 at very low luminosities, $L_{0.5-10\,{\rm keV}}\lesssim{\rm afew}\times10^{33}$~erg~s$^{-1}$, close to the typical threshold for the onset of the propeller regime \citep{Illiano_2025arXivJ17511, Bult:2019tr}.
In both systems, the unusually high pulse amplitudes, most likely due to favourable viewing geometries, allowed the detection of pulsations even beyond the expected accretion-propeller transition.
Given the relatively large pulse fraction observed in \maxi{} (see Sec.~\ref{sec:timing_analysis}), we explored whether accretion onto the neutron star in this system could also be taking place near, or perhaps just beyond, the propeller boundary.

Using the flux reported in Table~\ref{tab:params_spectrum}, we estimate that during the \xmm{} observation the source reached a $0.5-10$~keV luminosity of $\sim2\times10^{34}$~erg~s$^{-1}$ for an assumed distance of 5~kpc.
In accreting pulsars, accretion-powered pulsations can be sustained only if the magnetospheric truncation radius lies within the co-rotation radius, where the Keplerian frequency of the disk equals the stellar spin.
The latter is defined as $\mathrm{R_{co}} = [G M_{\mathrm{ns}}/(2 \pi \nu)^2]^{1/3} \simeq 1.68 \times 10^{6} \, \mathrm{M_{1.4}} \, \mathrm{P_{ms}}^{2/3} \, \mathrm{cm}$, where $\mathrm{M_{1.4}}$ is the NS mass ($M_{\mathrm{ns}}$) in units of 1.4 $\mathrm{M_\odot}$ and $\mathrm{P_{ms}}$ is the spin period in milliseconds. For a $1.4$~M$_\odot$ neutron star spinning at $P_{\rm ms}\simeq3.2$, this gives $R_{\rm co}\approx36$~km. To infer the mass accretion rate at the epoch of the \xmm{} observation, we extrapolated the unabsorbed $0.5-10$~keV flux to the $0.1-100$~keV band, obtaining $F_X=(7.4\pm0.2)\times10^{-12}$~erg~cm$^{-2}$~s$^{-1}$. The corresponding accretion rate is $\dot{M} = 4 \pi d^2 F_X R_{\mathrm{ns}} /(G M_{\mathrm{ns}}) \simeq 1 \times 10^{14} \, \mathrm{g \, s^{-1}}$ for a 1.4 M$_\odot$ NS with a $R_{\mathrm{ns}}=10$~km radius.
The magnetospheric (or truncation) radius, $R_{\mathrm{m}}$, can be expressed as a fraction $\xi$ of the Alfvén radius \citep[e.g.][]{Ghosh:1979aa}:
\begin{equation}
    R_{\mathrm{m}} = \xi  \left[ \frac{(B_{\mathrm{s}} R_{\mathrm{ns}}^3)^4}{(2 G M_{\mathrm{ns}} \dot{M}^2)} \right]^{1/7},
\end{equation}
where $B_{\mathrm{s}}$ is the NS surface magnetic field, and $G$ is the gravitational constant. Requiring $R_{\mathrm{m}} < R_{\mathrm{co}}$ for accretion to occur, we derived an upper limit on the magnetic field strength: 
\begin{equation}
    B_{\rm p}\ <\ 9\times10^{7}\ \mathrm{G}\ \Big(\frac{\xi}{0.5}\Big)^{-7/4}\Big(\frac{d}{5\,\mathrm{kpc}}\Big)\,k_{\rm bol}^{1/2}\Big(\frac{R_\mathrm{ns}}{10\,\mathrm{km}}\Big)^{-5/2}\Big(\frac{M_\mathrm{ns}}{1.4\,M_\odot}\Big)^{1/4},
\end{equation}
where $k_{\rm bol}$ represents the bolometric correction applied to $F_X$. Adopting $M=1.4\,\mathrm{M_\odot}$, $R=10$~km, and $\xi=0.5$ yields $B_{\rm p}\approx9\times10^{7}$~G. Including the distance uncertainty $d=(5\pm2)$~kpc \citep{Ravi:2017tl} gives $B_{\rm p}\approx(5.4\times10^{7}\text{-}1.3\times10^{8})$~G for $\xi=0.5$. Allowing for the typical range $\xi$ range 0.3-0.5 \citep[see, e.g., ][for more discussion on the parameter]{Burderi98a, Kulkarni:2013tf, Bozzo:2018aa, Campana_2018A&A} increases the limit by a factor  $(0.5/0.3)^{1.75}\simeq2.44$ at $\xi=0.3$, corresponding to $B_{\rm p}\approx2.2\times10^{8}$~G, and up to $\sim 3\times10^{8}$~G at 7~kpc. Conservatively, we therefore estimate the magnetic field in the range $B_{\rm p}\ \approx\ (0.5 - 3)\times10^{8}$~G for $d=5\pm2$~kpc, and $\xi=[0.3 - 0.5]$. 

It is worth noticing that this truncation-based bound is not consistent (smaller by a factor $\sim$3–10) with the independent estimate from the secular spin-down, which \rev{yields an upper limit on the polar dipole field of order $B_{\rm p}\lesssim 10^{9}$~G} if interpreted with a force-free torque law. Even adopting the most favourable geometry (i.e. a nearly aligned rotator, $\alpha = 0^\circ$), which increases the inferred dipole field by only a factor of $\sqrt{2}$, the value derived from the secular spin-down remains significantly higher than the truncation-based estimate. A possible way to reconcile the phenomenology is to relax the hard-barrier assumption at co-rotation: if accretion can persist in a mildly leaky propeller with $R_{\rm m}\gtrsim R_{\rm co}$, the truncation argument underestimates the true field \citep[see, e.g., the trapped-disc/weak-propeller framework of][]{DAngelo2012a, arnanson:2015aa}.

\rev{A similar phenomenology has recently been reported for the high-field Be/X-ray binary A~0538--66, where \nicer{} detected low-level pulsations at very low luminosity \citep{ducci:2025aa}. In that system, a partially open or “leaky” propeller was invoked to reconcile the magnetic field inferred from spin evolution with that required to allow residual accretion. Although A~0538--66 is a high-field Be/X-ray pulsar and therefore operates in a very different regime from \maxi{}, the idea that some inflowing matter can still cross $R_{\rm m}\gtrsim R_{\rm co}$ provides a useful parallel, supporting the plausibility of a mildly leaky propeller also in this source.}

An alternative (not mutually exclusive) possibility is that the secular $\dot\nu$ still carries a contribution from residual disc torques and/or phase–flux systematics, biasing the dipole estimate high. Longer time-baselines with additional outbursts will help to firm up the true long-term trend.

\subsection{Optical vs. X-ray emission}
    
\begin{figure}
    \centering
    \includegraphics[width=0.85\linewidth]{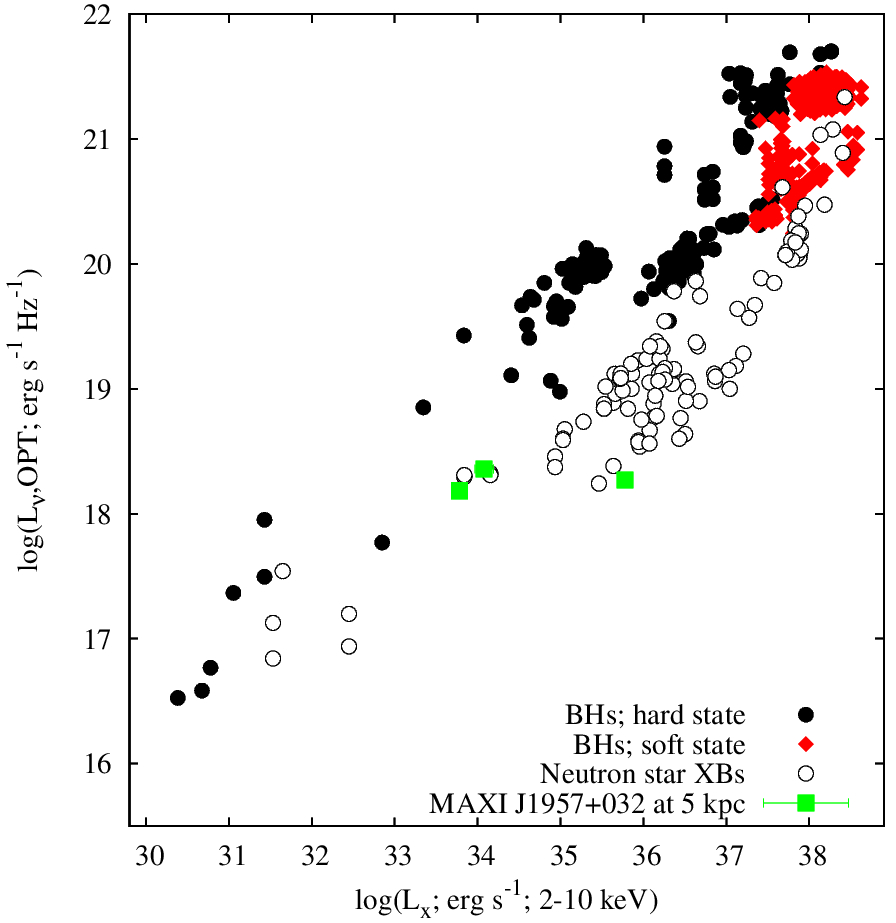}
    \caption{Optical luminosity (dereddened $\nu L_{\nu}$ in $g'$) versus X-ray luminosity (0.5--10\,keV) for \maxi{}. 
Black points and red diamonds show black-hole X-ray binaries in the hard and soft state, respectively; open circles represent neutron-star systems \citep{russel:2006aa, russel:2007aa}.}
    \label{fig:oir_x_corr}
\end{figure}

The optical emission in outbursting X-ray binaries empirically correlates with the X-ray luminosity, with distinct normalisations for neutron stars and black holes: at a given $L_X$, black holes are typically $\sim$an order of magnitude brighter in the optical/IR than neutron stars, and the global relation often follows $L_{\mathrm{OIR}} \propto L_X^{\,\beta}$ with $\beta \approx 0.5$--$0.7$, depending on band and accretion state \citep{russel:2006aa, russel:2007aa}. 
Using our dereddened $g'$ photometry (the same is also obtained using the $i'$ band one), matched to the nearest X-ray fluxes (\swift/XRT; \xmm{} for the final epoch), we place \maxi{} on the $L_{\mathrm{OIR}}$--$L_X$ diagram (Fig.~\ref{fig:oir_x_corr}). 

The source lies within the NS region, i.e.\ $\sim$1\,order of magnitude below the black-hole track at comparable $L_X$.
We note that the correlation for \maxi{} is not strictly positive throughout the campaign, which may be due to the source undergoing a state transition. 
Such transitions can modify the reprocessing geometry and jet contribution, thereby altering the OIR/X-ray coupling \citep[e.g.,][]{russel:2006aa}.

\rev{Figure~\ref{fig:SEDs_opt} shows that the spectral energy distribution on MJD~60805 is the brightest, consistent with the optical light curves in Fig.~\ref{fig:xray_opt_lc}, indicating that the optical peak likely occurred between MJD~60803.3 and MJD~60805.} The overall shapes are broadly consistent with irradiation-dominated disc emission (approximately flat to mildly red in $F_\nu$) as expected when X-ray reprocessing sets the optical output.

Interestingly, only the first SED (MJD~60803.1) shows an apparent excess at the longest wavelength, with the $i'$-band flux lying above the extrapolation from the bluer bands. This epoch coincides with the hardest X-ray spectrum (Fig.~\ref{Fig:spectral_evolution}), suggesting that the red excess could trace a contribution from a compact jet. In both black hole and neutron star transients, compact, continuously launched jets are characteristic of the hard accretion state and are quenched as the source softens \citep[e.g.,][]{Coriat2009, Baglio2018, Saikia2019}. The evolution of the SEDs in \maxi{}, from a relatively red first epoch to subsequently bluer spectra, is consistent with this picture, with the jet contribution fading as the X-rays soften. 

Among neutron star X-ray binaries, AMXPs are particularly prone to showing red or near-infrared excesses attributed to synchrotron emission from jets. This is likely due to their short orbital periods and relatively small accretion discs, which reduce the dominance of the thermal disc component, as well as their tendency to remain in hard states. Red excesses consistent with jet synchrotron emission have been reported in several AMXPs, including SAX~J1808.4$-$3658 \citep{Wang2001, Baglio2020}, IGR~J00291+5934 \citep{Lewis10}, XTE~J0929$-$314 \citep{Giles:2005tc}, and XTE~J1814$-$338 \citep{Krauss2005, Baglio2013}, among others. The red excess seen in \maxi{} during the first epoch may therefore represent a similar jet-related component.

\rev{Only upper limits are currently available in the radio band. The deepest one, from MeerKAT \citep{vandenEijnden2022}, corresponds to a radio luminosity of order a few $\times 10^{27}$~erg~s$^{-1}$ (for $d=5$~kpc). This is comparable to, or below, the faintest radio detections of compact jets in AMXPs such as SAX~J1808.4$-$3658 and IGR~J00291+5934, and therefore does not place strong constraints on the presence of a weak jet in \maxi{}. The tentative red excess we observe lies at the low end of the range of jet contributions inferred in those systems, but remains fully consistent with a relatively faint compact jet.}

Since reprocessing implies that optical photons arise from an extended region (likely the outer accretion disc), small X-ray/optical lags of order the light-travel time across the disc are expected. Delays of a few to several tens of seconds have indeed been measured in LMXBs hosting both neutron stars and black holes, and are interpreted as reprocessing signatures in the outer flow or on the donor star surface \citep[see, e.g.,][]{obrien:2002aa, munozdarias:2007aa, gandhi:2010aa, shahbaz:2023aa}. High-time-resolution optical monitoring, simultaneous with X-ray coverage, could reveal similar delays in \maxi{}, thereby constraining the size, geometry, and irradiation response of its disc.

The optical SEDs in Fig.~\ref{fig:SEDs_opt} (excluding the possible jet contribution in the first epoch) are roughly flat or slightly rising with frequency, consistent with emission from the outer disc. In standard disc models, such a shape is expected when part of the optical light is produced by the disc’s intrinsic thermal emission rather than purely by reprocessed X-rays. An optical brightening preceding the X-ray rise in the 2022 outburst \citep{Wang:2022us} suggests that the outer disc may start heating before the inner regions become X-ray bright. A plausible explanation for the delayed optical peak in the 2025 outburst is that, as the heating front propagates outward and the outer disc expands \citep[e.g.,][]{Dubus2001}, the emitting area grows, and the optical flux can continue to increase even while the X-rays already decay. Given the exceptionally short duration of this outburst, the disc expansion timescale may exceed the rapid X-ray decay timescale, naturally producing the observed delay. This sequence is qualitatively consistent with the thermal–viscous disc-instability framework commonly applied to transient systems \citep[see, e.g.,][]{Lasota01, Russell2019, Goodwin2020}.

\begin{acknowledgements}
We gratefully acknowledge Dan Bramich for his contributions to XB-NEWS and for helpful discussions. This work uses observations from the Las Cumbres Observatory Global Telescope Network. 
This work received financial support from INAF through the GRAWITA 2022 Large Program Grant.
A.B. acknowledges support through the European Space Agency (ESA) research fellowship programme. AM is supported by the National Spanish grant PGC2018-095512-BI00 (PI: Coti Zelati). This work was also partially supported by the program Unidad de Excelencia Maria de Maeztu CEX2020-001058-M. AP acknowledges support by INAF (Research Grant FANS and PULSE-X, PI: Papitto), the Italian Ministry of University and Research (PRIN MUR 2020, Grant 2020BRP57Z, GEMS, PI: Astone), and Fondazione Cariplo/Cassa Depositi e Prestiti (Grant 2023-2560, PI: Papitto).
\end{acknowledgements}

%
%

\bibliographystyle{aa} 
\bibliography{biblio2.bib}

\begin{appendix}
\onecolumn

\section{Swift/XRT observations and spectral results}

\begin{table}[h!]
\centering
\caption{\swift{}/XRT observations of \maxi{}.}
\label{tab:swift_obs}
\begin{tabular}{lccc}
\hline\hline
ObsID & Mode & Start Time (UT) & Exposure (s)\\
\hline
00019768001 & PC & 2025-05-06T21:54:15 & 1135.8\\
00033770034 & WT & 2025-05-08T03:50:55 & 1207.2\\
00033770035 & WT & 2025-05-09T08:04:56 & 949.4\\
00033770037 & WT & 2025-05-09T13:52:56 & 1879.4\\
00033770038 & WT & 2025-05-10T02:46:56 & 1014.7\\
00033770039 & WT & 2025-05-10T09:03:56 & 871.9\\
00033770040 & WT & 2025-05-10T14:59:16 & 1031.8\\
00033770041 & WT & 2025-05-10T21:15:56 & 1051.6\\
00033770042 & WT & 2025-05-11T03:33:56 & 1019.6\\
03400042001 & PC & 2025-05-15T00:59:56 & 1836.4\\
\hline
\end{tabular}
\end{table}

\begin{table}[h!]
\centering
\caption{Best-fitting model continuum for each \swift{}/XRT observation of \maxi{}.}
\label{Table:spectra_Swift}
\begin{tabular}{lcccccc}
\hline\hline
ObsID & $N_{\mathrm{H}}$ & $\Gamma$ & $F_{0.5-10}$ & $F_{0.5-2}$ & $F_{2-10}$ & $\chi^{2}$/d.o.f.\\
 & ($10^{21}$ cm$^{-2}$) & & ($10^{-10}$ erg cm$^{-2}$ s$^{-1}$) & ($10^{-10}$ erg cm$^{-2}$ s$^{-1}$) & ($10^{-10}$ erg cm$^{-2}$ s$^{-1}$) & \\
\hline
00019768001 & $1.1\pm0.2$ & $1.66^{+0.07}_{-0.06}$ & $5.1\pm0.2$ & $1.8\pm0.1$ & $3.4\pm0.2$ & 92.23/93\\
00033770034 & $1.8\pm0.1$ & $1.85\pm0.03$ & $3.40\pm0.05$ & $1.39\pm0.04$ & $2.01\pm0.05$ & 263.99/242\\
00033770035 & $5\pm1$ & $2.8\pm0.2$ & $0.51^{+0.08}_{-0.06}$ & $0.37^{+0.09}_{-0.07}$ & $0.14\pm0.02$ & 20.26/24\\
00033770037 & $3.3\pm0.5$ & $2.7\pm0.1$ & $0.45^{+0.04}_{-0.03}$ & $0.32^{+0.04}_{-0.03}$ & $0.14\pm0.01$ & 54.19/50\\
00033770038 & $4.2^{+0.9}_{-0.8}$ & $2.8\pm0.2$ & $0.34^{+0.05}_{-0.04}$ & $0.25^{+0.06}_{-0.04}$ & $0.09\pm0.01$ & 13.65/20\\
00033770039$^{(\star)}$ & $7.2^{+2.9}_{-2.4}$ & $3.2\pm0.5$ & $0.2\pm0.1$ & $0.2\pm0.1$ & $0.04\pm0.01$ & --\\
00033770040$^{(\star)}$ & $2.1^{+1.6}_{-1.3}$ & $2.9^{+0.5}_{-0.4}$ & $0.06^{+0.02}_{-0.01}$ & $0.05^{+0.03}_{-0.01}$ & $0.015^{+0.005}_{-0.004}$ & --\\
\hline
\end{tabular}
\tablefoot{
$N_{\mathrm{H}}$ is the absorption column density, and $\Gamma$ is the photon index.
$F_{0.5-10}$, $F_{0.5-2}$, and $F_{2-10}$ are the unabsorbed fluxes estimated in the 0.5–10, 0.5–2, and 2–10 keV energy bands.
$^{(\star)}$ Spectra grouped with 10 counts/bin and fitted using the Cash statistic (Sect.~\ref{sec:swift_data}).
}
\end{table}

\section{Optical photometry and calibration}\label{app:optical}

During the 2025 outburst, \maxi{} was clearly detected in the $g'$, $r'$, and $i'$ bands between MJD~60803 and MJD~60806.
Images obtained in the $Y$ filter yielded only non-detections or upper limits.  
Photometry was extracted with the \textsc{XB-NEWS} pipeline \citep{Russell2019, Goodwin2020}, which performs multi-aperture photometry on \textsc{SExtractor}-detected sources \citep{Bertin1996} and calibrates instrumental magnitudes against the ATLAS-REFCAT2 catalogue \citep{Tonry2018}, supplemented with Pan-STARRS~DR1 and APASS~DR10 \citep{Chambers2016, Henden2018}.  
A spatially variable photometric model \citep{Bramich2012} provided zeropoints and PSF corrections through an iterative, outlier-weighted fit.  
Colour terms were neglected due to limited multi-band overlap, introducing systematics below 2\%.  
Measurements with uncertainties $>0.25$~mag were excluded.  
The final magnitudes, uncorrected for reddening, are listed in Table~\ref{tab:optical_photometry}.  

\begin{table}[h!]
\renewcommand{\arraystretch}{1.15}
\centering
\caption{Optical photometry of \maxi{} during the 2022 and 2025 outbursts.}
\label{tab:optical_photometry}
\begin{tabular}{lccc}
\hline\hline
Filter & MJD & Magnitude & Telescope site\\
\hline
$g'$ & 59750.56490 & $19.087\pm0.016$ & coj\\
     & 59751.09793 & $19.126\pm0.027$ & tfn\\
     & 59751.37698 & $19.321\pm0.049$ & lsc\\
     & 60803.10129 & $19.754\pm0.019$ & cpt\\
     & 60803.30945 & $19.614\pm0.024$ & lsc\\
     & 60805.30250 & $19.433\pm0.033$ & lsc\\
     & 60806.17103 & $19.684\pm0.056$ & tfn\\
\hline
$r'$ & 59750.57027 & $19.132\pm0.021$ & coj\\
     & 59751.10336 & $19.026\pm0.032$ & tfn\\
     & 59751.38242 & $18.923\pm0.086$ & lsc\\
     & 60803.09805 & $19.765\pm0.036$ & cpt\\
     & 60803.30620 & $19.501\pm0.033$ & lsc\\
     & 60805.29926 & $19.362\pm0.036$ & lsc\\
     & 60806.16770 & $19.507\pm0.055$ & tfn\\
\hline
$i'$ & 59750.56759 & $18.943\pm0.034$ & coj\\
     & 59751.10062 & $18.997\pm0.041$ & tfn\\
     & 59751.37965 & $19.123\pm0.079$ & lsc\\
     & 60803.09538 & $19.482\pm0.049$ & cpt\\
     & 60803.30353 & $19.487\pm0.048$ & lsc\\
     & 60805.29658 & $19.261\pm0.044$ & lsc\\
     & 60806.16499 & $19.693\pm0.068    $ & tfn\\
\hline
\end{tabular}
\tablefoot{Telescope site codes: coj = Siding Spring Observatory; lsc = Cerro Tololo; tfn = Teide Observatory; cpt = Sutherland. Magnitudes are uncorrected for reddening ($A_{g'}=0.04\pm0.01$~mag, $A_{r'}=0.03\pm0.01$~mag, $A_{i'}=0.02\pm0.01$~mag).}
\end{table}

\begin{figure}[h!]
\centering
\includegraphics[width=0.75\linewidth]{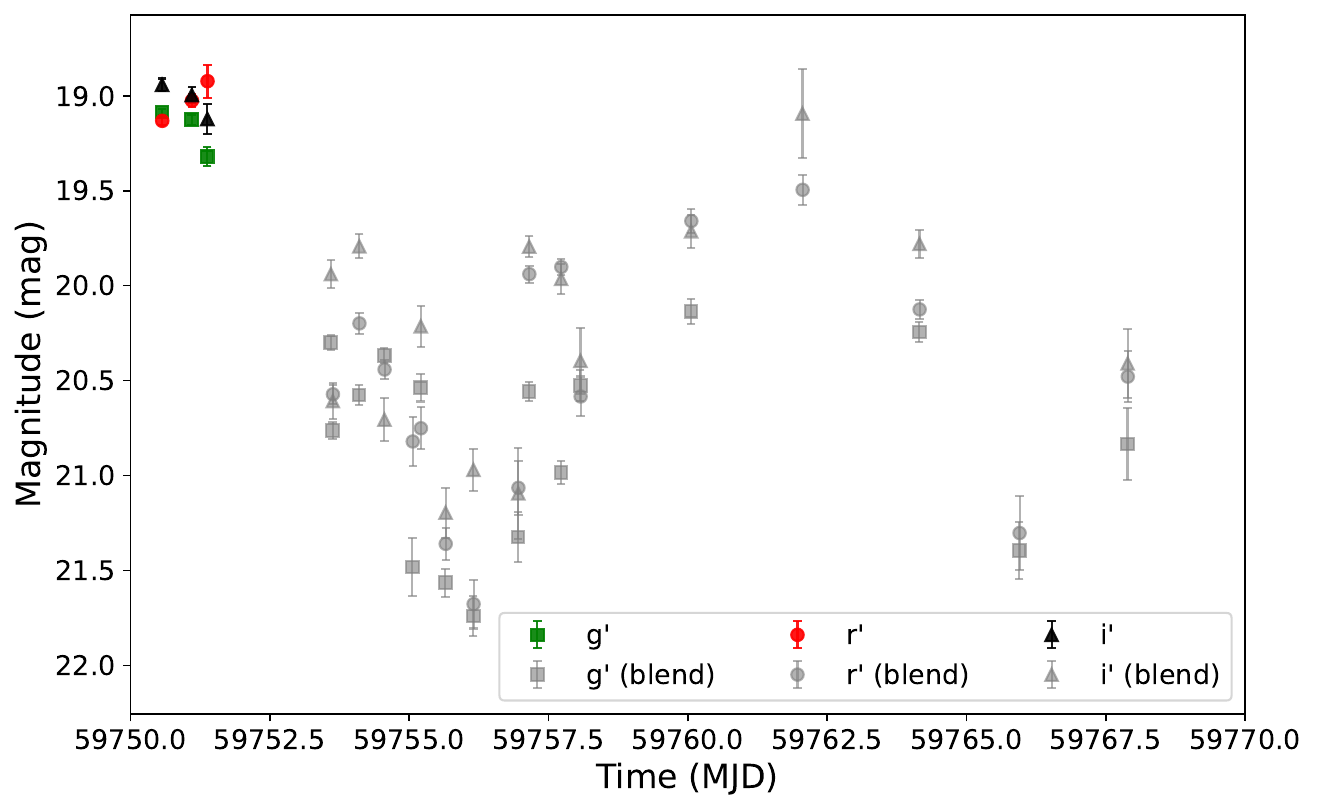}
\caption{Optical light curves of \maxi{} obtained with LCO between 2022 June 20 (MJD~59750) and 2022 July 10 (MJD~59770), covering the entire outburst. Grey points mark blended measurements with the nearby star. The secondary maximum is clearly visible and fainter than the primary.}
\label{fig:2022outburst}
\end{figure}

\end{appendix}

\end{document}